%% file: main.tex
\documentclass[twocolumn,english,aps,pra,superscriptaddress,amsmath,amssymb,floatfix,nofootinbib,longbibliography
]{revtex4-2}

\usepackage{amsthm}
\usepackage{amsfonts}
\usepackage{siunitx}
\usepackage{amsmath}
\usepackage{amssymb}
\usepackage{graphicx}
\usepackage{verbatim}
\usepackage[colorlinks]{hyperref}
\usepackage{tikz}
\usepackage{pgfplots}
\usepackage{adjustbox}
\usepackage{braket}
\usepackage{xcolor}
\usepackage{physics}
\usepackage{amssymb} 
\usepackage{graphicx}
\usepackage{dcolumn}
\usepackage{bm}
\usepackage{mathtools}
\usepackage{hyperref}
\usepackage{mathrsfs}
\usepackage{dashrule}
\usepackage{caption}
\usepackage{subcaption}
\usepackage{quantikz}
\usepackage[version=4,arrows=pgf-filled,
textfontname=sffamily,
mathfontname=mathsf]{mhchem}

\usepackage[font=small,labelfont=bf,
   justification=justified,
   format=plain]{caption}

\definecolor{linkcolor}{RGB}{0,83,166}
\hypersetup{
  colorlinks = true,
  allcolors = {linkcolor}
}

\begin{document}

\title{Effects of Trotter Error, Digitization Error, and Initial State Overlap on Tapered Quantum Phase Estimation for Minimum Eigenvalue Computation}

\author{Elijah Pelofske}
\email[]{epelofske@lanl.gov}
\affiliation{Quantum \& Condensed Matter Physics, Los Alamos National Laboratory, Los Alamos, NM, USA}
\affiliation{Center for Quantum Computing, Los Alamos National Laboratory, Los Alamos, NM, USA}

\author{Stephan Eidenbenz}
\email[]{eidenben@lanl.gov}
\affiliation{Information Sciences, Los Alamos National Laboratory, Los Alamos, NM, USA}
\affiliation{Center for Quantum Computing, Los Alamos National Laboratory, Los Alamos, NM, USA}

\begin{abstract}

\input{main_abstract.tex}

\end{abstract}

\maketitle

\input{main_text}

\clearpage

\bibliographystyle{apsrev4-2-titles}
\bibliography{references}
\end{document}

%% file: main_abstract.tex
We numerically implement full quantum circuits of optimal tapered Quantum Phase Estimation (in the form of an implementable approximate bandwidth-limited DPSS taper) for the purpose of minimum eigenvalue computation, where the Hamiltonian time-evolution unitary is implemented using standard Trotterization, and the taper state is constructed using an optimized bandwidth-limited taper. 
We illustrate, on representative quantum Hamiltonians, with tractable exact classical numerical quantum circuit simulations, how tapered QPE performs with respect to optimal phase sampling rate, absolute error, Shannon entropy of the optimal phase distribution, when the tQPE algorithm parameters are changed. 
Those algorithm parameters are Trotter error, finite sampling, imperfect initial states, total evolution time, and digitization error (number of phase qubits). 
The reported numerical experiments include up to $\ell=10$ phase qubits of precision with $m=3$ additional phase-register error suppression qubits, on a $4$-qubit Heisenberg quantum magnet model Hamiltonian, using up to 12th-order Trotterization. 
We numerically show i) the steady-state optimal phase sampling rate is determined by the initial state overlap with the ground-state, ii) with respect to phase sampling probability, or overall eigenvalue sampling error rate, there is no strong evolution time dependence for tQPE unless there is high Trotter error, iii) the approximated time-evolution unitaries in tQPE, like in standard QPE, can result in substantial ringing, which leads to non-physical eigenvalue estimates.

%% file: main_text.tex
%%%%%%%%%%%%%%%%%%%%%%%%%%%%%%%%%%%%%%%%%%%%%%%%
\section{Introduction}\label{section:Introduction}
%%%%%%%%%%%%%%%%%%%%%%%%%%%%%%%%%%%%%%%%%%%%%%%%

Quantum phase estimation (QPE) is arguably the most powerful quantum algorithm as it enables us to compute minimum eigenvalues of exponentially large matrices such as quantum Hamiltonians~\cite{nielsen2010quantum, kitaev1995quantummeasurementsabelianstabilizer, Abrams_1999, Dob_ek_2007}. Such minimum eigenvalue calculations are important in a wide range of applications, including quantum chemistry, strongly correlated quantum materials, and condensed matter physics, to name just a few~\cite{aspuru2005simulated, whitfield2011simulation, PhysRevA.79.062314, j26w-w48j, Kivlichan_2020, PhysRevX.6.031007}. Moreover, QPE is also a core sub-routine in many quantum algorithms including HHL~\cite{PhysRevLett.103.150502} and Shor's algorithm~\cite{365700, Shor_1997}. QPE has certain implementation considerations, namely non-physical ``ringing'' (spectral leakage) which introduces unwanted errors, even in the idealized setting of high-overlap initial states and no Trotter error. To this end, various QPE algorithmic advances have been proposed~\cite{Rall_2021, Mande_2026, nagaj2009fastamplificationqma, jiang2024quantummetropolissamplingweak}, the most effective of which is a type of optimal tapered QPE that uses an optimal Discrete Prolate Spheroidal Sequences (DPSS) state preparation in order to exponentially suppress ringing and thereby non-physical error rates~\cite{Patel_2026}, which is especially relevant for accurate minimum eigenvalue computation. QPE, and tQPE in particular, has a wide range of potential applications on future quantum computers. Here, we examine the task of minimum eigenvalue computation of quantum Hamiltonians.

In this study, we numerically implement, end-to-end, trotterized digital quantum circuits of the optimal coherent tapered QPE (tQPE) algorithm described in Ref.~\cite{Patel_2026}, on small quantum Hamiltonians thereby allowing full (classical) statevector simulations of these circuits. Moreover, we numerically highlight the exact mechanism by which tQPE results in lower error rates. We examine the effects of Trotter error, imperfect and perfect initial states, evolution time, and digitization error. In this way, we are emulating what a large-scale quantum computer would do in order to execute tapered QPE, with the goal of ground-state computation of the minimum eigenvalue of a quantum Hamiltonian. The error rate of tQPE for computing a minimum eigenvalue is determined by a combination of three factors: 

\begin{enumerate}
    \item Initial state overlap with the ground-state. 
    \item How well the controlled unitaries which approximate $e^{i H t}$ are implemented (we will refer to this as \emph{Trotter error}). 
    \item Digitization error, determined by the number of bits of precision used in the phase register. 
\end{enumerate}

In this study we present numerical executions of tQPE which explain how these three factors influence the error rate of tQPE for minimum eigenvalue computation. 
Note that with standard QPE, e.g. not the optimal taper QPE of Ref.~\cite{Patel_2026}, there is a fourth factor which is the evolution time $t$ that is used in $e^{i H t}$~\cite{pelofske2026numericalexperimentsparametersetting}.

\begin{figure*}[ht!]
    \centering
    \includegraphics[width=0.999\linewidth]{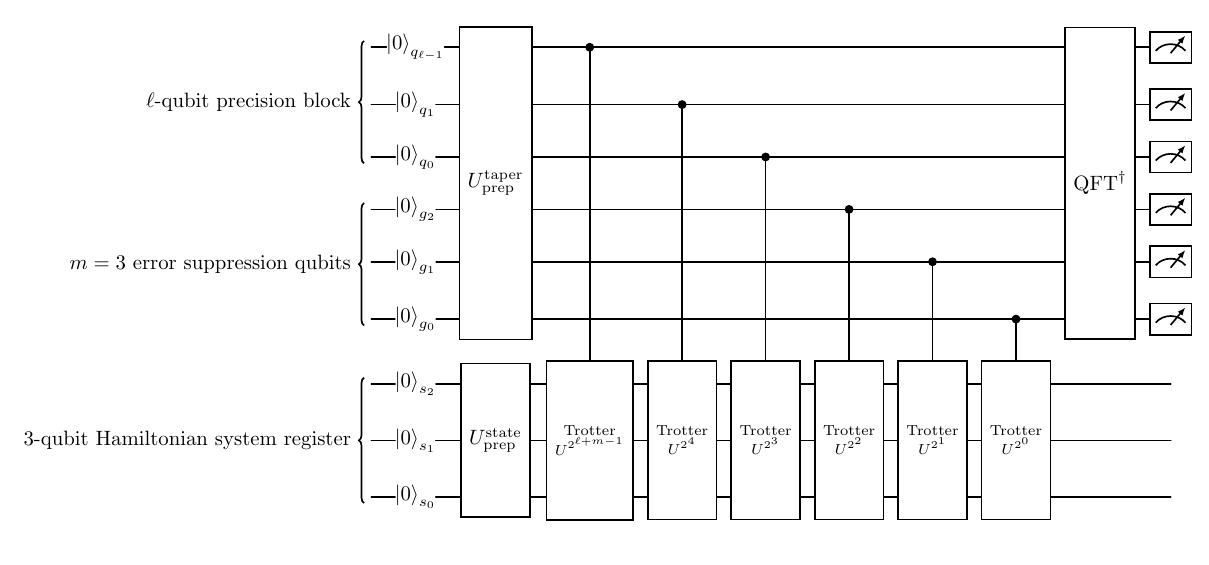}
    \caption{\textbf{Tapered QPE circuit diagram.} The qubit registers are divided into three regions; the first is the $\ell$ register which is the resolution of the phase estimate, next is the $m$ register which is part of the ancilla phase-qubit register. However, this $m$-qubit register specifically suppresses ringing (spectral leakage) errors, and does not directly contribute to the precision of the digitized phase readout. The third is the system Hamiltonian register, where the Hamiltonian is evolved using trotterization, with the system having been initialized in a state that has high overlap with the ground-state of the quantum Hamiltonian $\mathcal{H}$ we are simulating. Here, we set $\ell=3$, $m=3$, and $n=3$, although in practice all three of these qubit registers can be set to any size $\geq 1$, with the only implications being overall circuit or construction complexity. At the conclusion of the circuit, all phase-register qubits ($m+\ell$) are measured in the computational $Z$ basis, and in this case the qubits in the Hamiltonian evolution register are not measured. As with standard QPE, the time evolution for each subsequent least significant bit requires a power of 2 more total evolution time for that phase qubit register, which makes having many phase (ancilla) qubits very costly in terms of total single and two-qubit gate count. 
    The tQPE circuit is implemented such that each trotterized unitary $\Tilde{U}$ is constructed using the same trotter order $p$ and trotter steps $r$ to approximate the full Hamiltonian time-evolution unitary of $e^{i H t}$, where $t=t_0\cdot 2^j$ and $j \in \{0, \ldots, \ell+m-1\}$. 
    }
    \label{fig:tapered_QPE_circuit}
\end{figure*}

%%%%%%%%%%%%%%%%%%%%%%%%%%%%%%%%%%%%%%%%%%%%%%%%
\section{Methods}\label{section:Methods}
%%%%%%%%%%%%%%%%%%%%%%%%%%%%%%%%%%%%%%%%%%%%%%%%

We begin by outlining the tQPE algorithm, and the terminology and notation that we will use. Our goal is to compute the minimum eigenvalue, with as much precision as we are able to, of a quantum Hamiltonian of interest $\mathcal{H}$ with $n$ qubits. 
Just as in standard ``textbook'' QPE, the core computation is implementing the time dynamics computation of $e^{i H t}$, in the form of a controlled unitary time evolution gate. We will use standard trotterization to approximate $e^{i H t}$, using progressively higher Trotter orders (denoted as $p$) and more Trotter steps (denoted as $r$)~\cite{trotter1959product, suzuki1976generalized, suzuki1985decomposition, Hatano_2005, lopezcerezo2025rigorousintroductionhamiltoniansimulation, Berry_2006}. The standard Trotter-Suzuki implementation, in Qiskit~\cite{javadiabhari2024quantumcomputingqiskit, matthew_treinish_2026_18992040} (version \texttt{2.3.1}), uses a recursive construction, which results in extremely high-depth and high gate-count circuits. Larger Trotter order, $p$, more dramatically decrease simulation error, at the cost of much larger quantum circuits, whereas Trotter steps $r$ decreases simulation error more gradually. We use these methods because they are standard methods, with known error-scaling properties. However, many other improved product formula methods exist in various contexts that would reduce the absolute gate counts in a tQPE implementation~\cite{maležič2026efficienttrottersuzukischemeslongtime, maležič2026reducinggatecountefficient, Ostmeyer_2023, Gibbs_2025, gibbs2025learningcircuitsinfinitetensor, Childs_2019, Tranter_2019, hastings2014improvingquantumalgorithmsquantum}, which we leave for future work. We implement QPE circuits on quantum Hamiltonians with a small number of qubits, e.g. $n=4$, which makes the numerical simulation of these circuits tractable (parallelized with High Performance Computing). This also means that the quantum Hamiltonian can be diagonalized exactly, which we implement using the Python 3 library NumPy~\cite{harris2020array} that in turn relies on the LAPACK linear-algebra library~\cite{129995}. 

\begin{figure}[ht]
    \centering
    \includegraphics[width=0.999\linewidth]{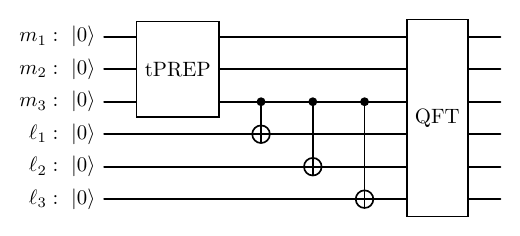}
    \caption{\textbf{Approximate taper state construction, using the example $m=\ell=3$.} This is the recipe described for preparing an approximate version of the taper unitary in Ref.~\cite{Patel_2026}, notated as $U_{\mathrm{prep}}^{\mathrm{taper}}$ in Fig.~\ref{fig:tapered_QPE_circuit}.  
    }
    \label{fig:taper_state_vector_circuit}
\end{figure}

The quantum Hamiltonian we consider is a disordered $\pm J$ fully-connected Heisenberg spin glass model, 

\begin{equation}
\label{eq:heisenberg_glass}
\mathcal{H}
\,=\,
\sum_{i<j} 
\Big(
  J^{x}_{ij}\,\sigma_i^{x}\sigma_j^{x}
 +J^{y}_{ij}\,\sigma_i^{y}\sigma_j^{y}
 +J^{z}_{ij}\,\sigma_i^{z}\sigma_j^{z}
\Big).
\end{equation}

This model serves as a good test-case Hamiltonian because the ground-state can be entangled and degenerate, and the model itself is highly frustrated particularly at large system sizes. Each $J_{ij} \in \{+1, -1\}$ is drawn randomly and independently for each coupling, making this a type of anisotropic disordered quantum magnet~\cite{Itoi_2024, Georges_2000} similar in spirit to the disorder of the Sherrington-Kirkpatrick models~\cite{PhysRevLett.35.1792, de_Almeida_1978}. We will focus on circuit implementation, and executions, of tQPE on a single $4$ qubit instantiation of this Heisenberg quantum Hamiltonian model, and a single $3$ qubit instance of this model. Appendix~\ref{section:appendix_model_values} supplies the exact model weights. The particular Hamiltonian instance does not change the core algorithmic characteristics of tQPE that we examine, and although larger quantum Hamiltonians could be implemented, the high classical statevector compute cost makes this infeasible to do for much larger Hamiltonians. We focus on single Hamiltonian instances as representative examples to illustrate the properties of tQPE. Moreover, the properties of tQPE do not substantially change across different problem instances; the primary thing that changes is the overlap that different initial states have with the Hamiltonian ground-state.  

\begin{figure*}[ht!]
    \centering
    \includegraphics[width=0.49\linewidth]{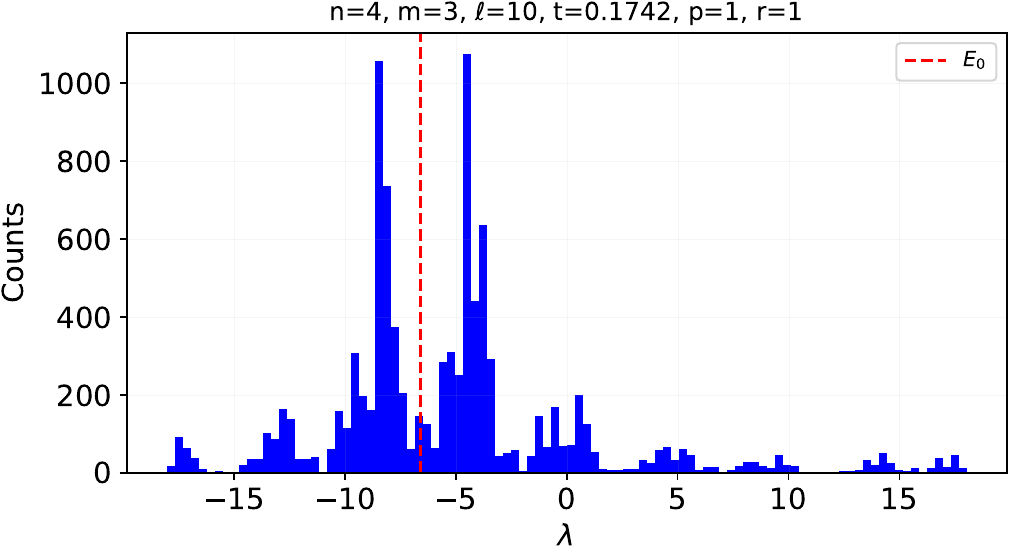}
    \includegraphics[width=0.49\linewidth]{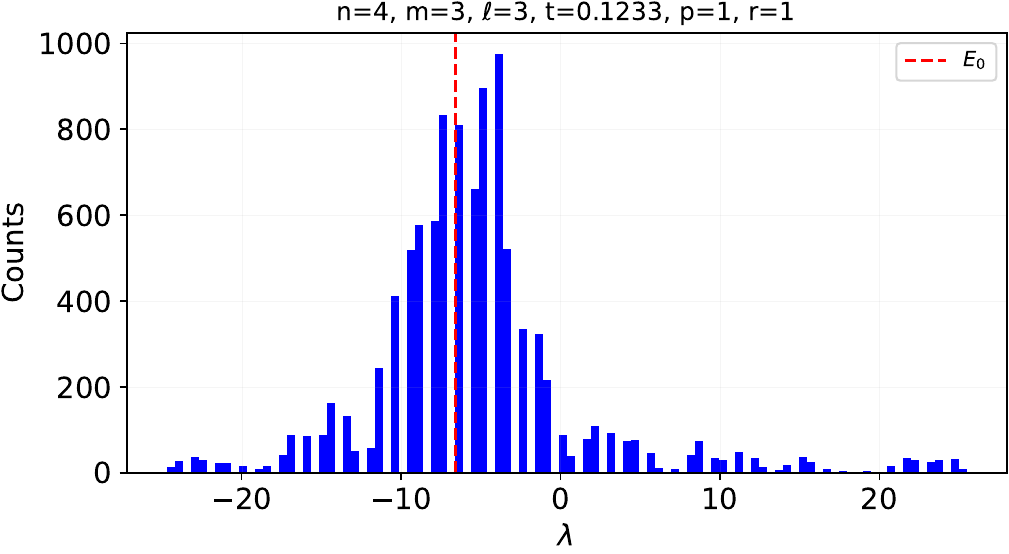}
    \includegraphics[width=0.49\linewidth]{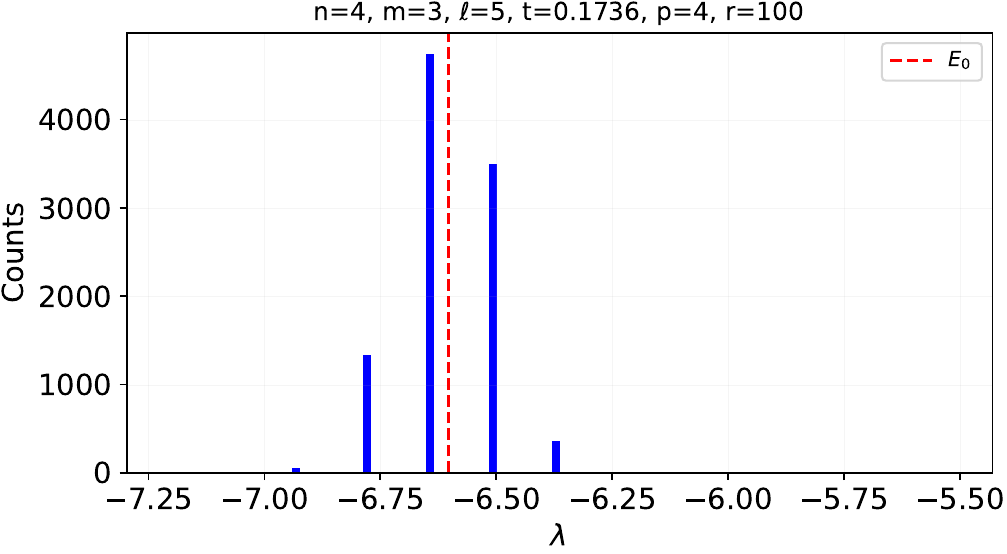}
    \includegraphics[width=0.49\linewidth]{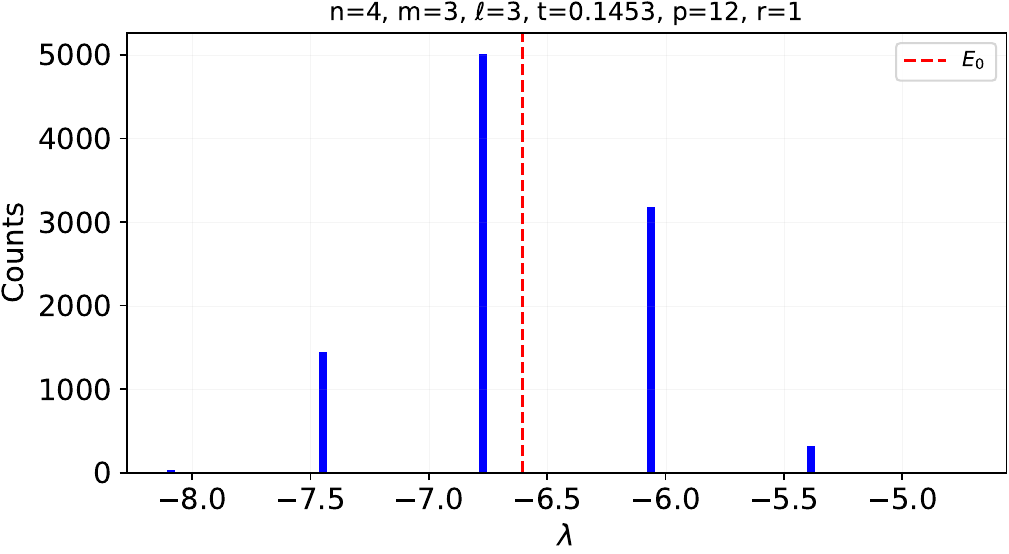}
    \includegraphics[width=0.49\linewidth]{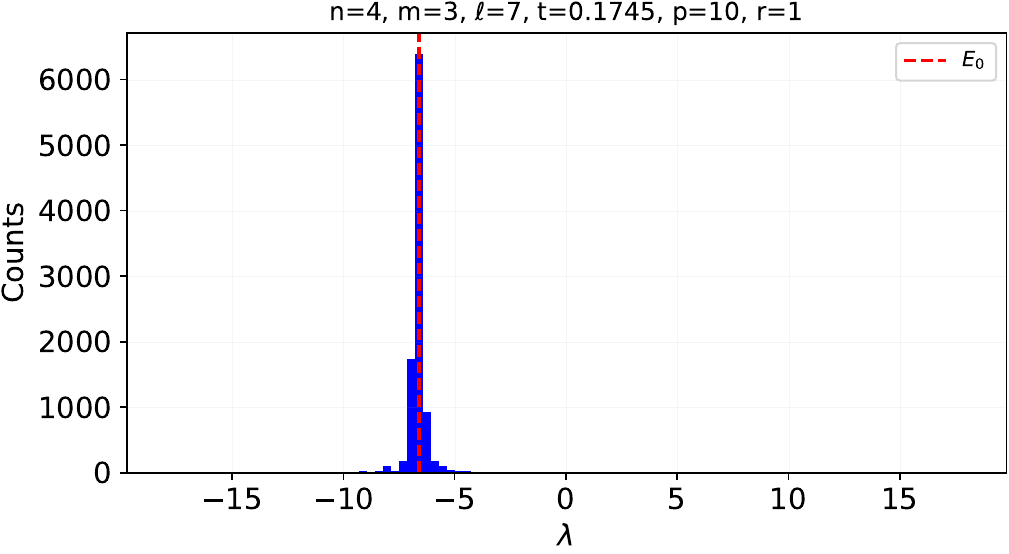}
    \includegraphics[width=0.49\linewidth]{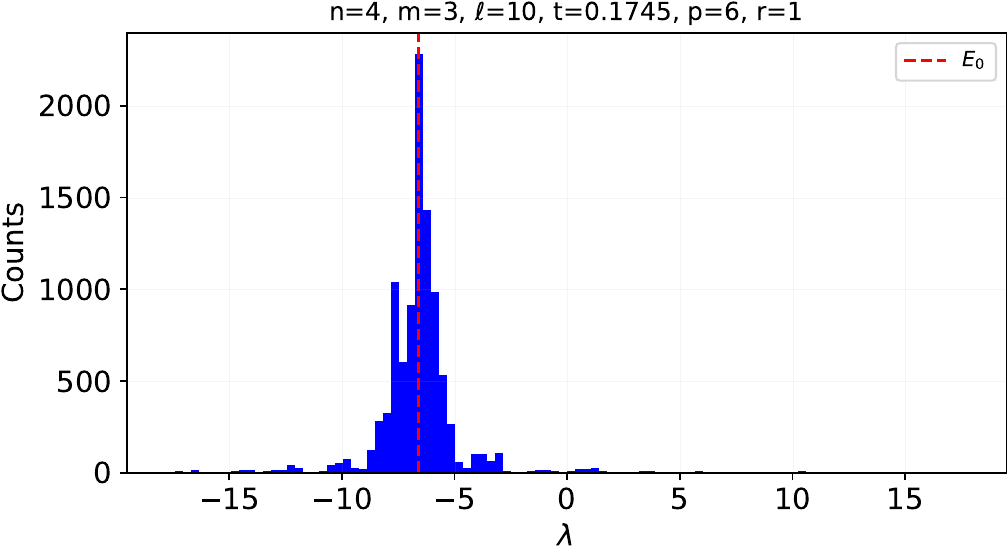}
    \caption{\textbf{Eigenvalue estimate distributions from tQPE, with a perfect initial state.} The true minimum eigenvalue is marked with the vertical red line as $E_0$. The distribution of eigenvalue estimates, $\lambda$, is computed from the distribution of $10^4$ sampled phase qubit measurements and the evolution time bound given by Eq.~\eqref{equation:evolution_time_bound}. Note that the decoded eigenvalue estimates $\lambda$ are not necessarily physical to the Hamiltonian, in fact when the Trotter error is very high the eigenvalue estimates can extend well below the true minimum eigenvalue. 
    \textbf{The top-row} shows two examples where the Trotter error is very high, where the time evolution was performed using only $p=r=1$. 
    \textbf{The middle-row} shows two examples where the Trotter error is sufficiently small such that the prominence of the $\Phi^*$ set of phases (equal to $7$ because $m=3$) is clearly visible. This middle row shows the absolute best-case tQPE sampling, where the Trotter error is effectively negligible. Notice that the phases for $\ell=5$ concentrate closer to $E_0$ compared to $\ell=3$. 
    \textbf{The bottom-row} shows examples where $\ell$ is large and the Trotter error is sufficiently large such that the most likely eigenvalue is a good approximation of $E_0$, but the distribution still has wide variance. 
    }
    \label{fig:eigenvalue_distributions_perfect_init}
\end{figure*}

QPE is a very general quantum algorithm, but here we focus on the use-case of QPE computing the minimum eigenvalue of a quantum Hamiltonian. To this end, QPE computes an estimate of the phase $\phi$, which corresponds to a digitization of the minimum eigenvalue of the quantum Hamiltonian $E_0$. 

The parameters that determine the minimum eigenvalue phase accuracy in tQPE are $\ell$ and $m$. Parameter $m$ determines the level of error-suppression, in terms of reducing ringing that is outside a narrow window of phase. Parameter $\ell$ then determines the digitization accuracy of the phase, and the actual bit-precision estimate of the phase is determined by $\ell+m$ (in the literal sense of how many bits are measured).

Fig.~\ref{fig:tapered_QPE_circuit} renders an example tQPE circuit. The key difference of the circuit, relative to standard QPE~\cite{pelofske2026numericalexperimentsparametersetting}, is that the state preparation of the phase qubit registers instead of being all Hadamard gates, is now a more complex state. Fig.~\ref{fig:taper_state_vector_circuit} defines the approximate construction of the optimal coherent tapered state, developed in ref.~\cite{Patel_2026}, with more numerical details given in Appendix~\ref{section:appendix_taper_states}. Importantly, here our numerical implementation uses the \emph{approximate} taper state construction described in Ref.~\cite{Patel_2026}, for the purposes of making the numerical implementation of the tQPE circuit reasonable. In particular, the taper state could create an exact DPSS taper; however this could be quite costly to implement~\cite{Patel_2026}. Therefore, Ref.~\cite{Patel_2026} introduced optimized bandwidth-limited DPSS taper states that are efficient to prepare, similar to truncated DPSS taper states. Importantly, although approximate, these bandwidth-limited taper DPSS taper states have near optimal performance compared to the exact DPSS taper~\cite{Patel_2026}. Therefore, we will generally refer to this taper QPE implementation as ``optimal coherent taper QPE'', following Ref.~\cite{Patel_2026}, however the taper states used here are approximate.

The total number of phase qubits used is $\ell+m$, where $\ell$ determines the effective digitization precision of the minimum eigenvalue. For the numerical experiments performed in this study, and as Fig.~\ref{fig:tapered_QPE_circuit} shows, we use $m=3$, which corresponds to an error of $\approx 10^{-8}$~\cite{Patel_2026} (which corresponds to the probability of sampling a phase outside of the high confidence region of tQPE, not an absolute error), which is more than sufficient for the problem size and number of phase qubits we consider. For this study, we use the bandwidth-limited \texttt{tPREP} taper vectors for $\ell=3$ up to $\ell=10$, with $m=3$ fixed. Each of these vectors changes slightly as $\ell$ increases, but not substantially~\cite{Patel_2026}, see Appendix~\ref{section:appendix_taper_states}.

In the idealized setting of very low (effectively zero) Trotter error, perfect initial state, and perfectly chosen evolution time for standard QPE~\cite{pelofske2026numericalexperimentsparametersetting, nielsen2010quantum, kitaev1995quantummeasurementsabelianstabilizer}, there is an important distinction between tQPE and standard QPE. Unlike standard QPE~\cite{pelofske2026numericalexperimentsparametersetting, nielsen2010quantum, kitaev1995quantummeasurementsabelianstabilizer}, tQPE does not concentrate the measured phase probability into the nearest digitized phase, instead, tQPE concentrates the phase measurements into a set of phase bitstrings. The number of optimal phase bitstrings in tQPE~\cite{Patel_2026}\footnote{Not considering a pathological case where the true phase is exactly between two phase ``grid-points''.} is $2K+1$, where $K = 2^{m-1}-1$. The guarantee that tQPE gives is that when $m$ is sufficiently large (and $m=3$ is sufficiently large for practical uses of QPE), the probability that measured phase falls outside of this set of $2K+1$ bitstrings is exponentially small~\cite{Patel_2026}. In this way, the digitization error in tQPE is slightly different to standard QPE~\cite{pelofske2026numericalexperimentsparametersetting} in the sense that the digitization error is now determined by a set of $2K+1$ bitstrings, instead of a single phase bitstring. When $\ell$ is large, this sub-space of $2K+1$ optimal phase qubit readouts becomes very close to each other. We will denote the set of tQPE ``optimal'' phases $\phi$ as $\Phi^*$, and the size of $\Phi^*$ is exactly $2^m-1$ for the taper construction of Ref.~\cite{Patel_2026}. We refer to this set as the ``optimal'' phases because these are the phases that we would ideally want to measure in a tQPE computation if there was negligible trotter error, and our initial state overlap was high. Because we focus on $m=3$, we will then increase $\ell$ incrementally and empirically measure the accuracy of tQPE as $\ell$ increases. When $m=3$, the number of phases in the high confidence region is $7$ which means that there is a single phase that is the nearest digitization to the true phase, and then there are 3 phases immediately adjacent to that optimal digitized phase on either side. Note that the phases in $\Phi^*$ are determined by $\ell, m$, the quantum Hamiltonian we are simulating, and the total evolution time $t$ in $e^{i H t}$.

tQPE is designed to produce effectively no ``side-lobes'' (also referred to as ``ringing''), in the sense of phase probabilities that fall outside of this intended range of optimal phases. However, in the practical implementation of tQPE we use an approximation of $U = e^{i H t}$, which we denote as $\Tilde{U}$. One of the properties that we numerically characterize is non-physical eigenvalue estimates, a form of ``ringing'', due to the approximation of each $\Tilde{U}_j$. All circuit descriptions are implemented in Qiskit~\cite{javadiabhari2024quantumcomputingqiskit} and decomposed down to a universal gateset of single and two qubit gates. Each parameter combination, instantiated as a quantum circuit, is sampled using exactly $10^4$ shots with the statevector simulator in \texttt{Qiskit Aer}~\cite{javadiabhari2024quantumcomputingqiskit}, with no error or noise model applied. 

\begin{figure*}[ht]
    \centering
    \includegraphics[width=0.49\linewidth]{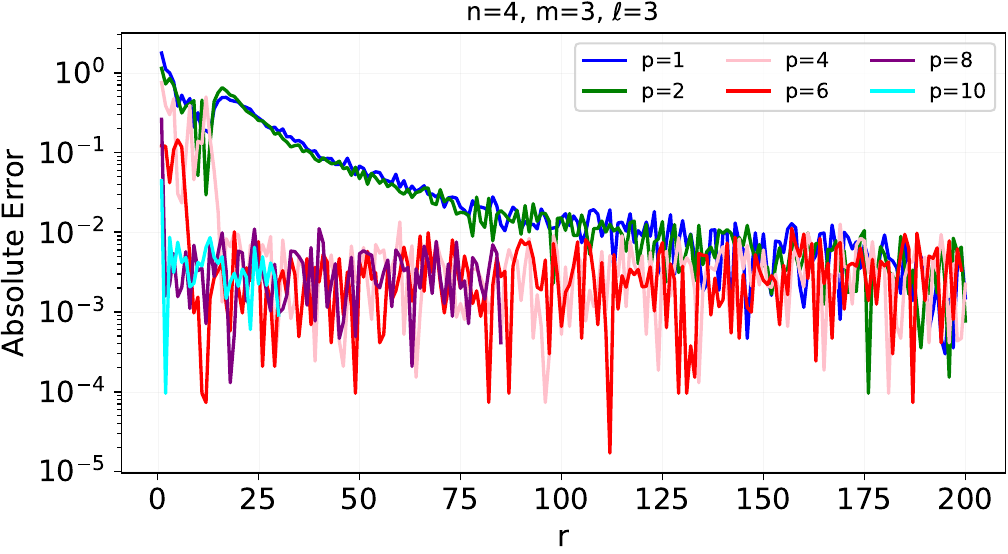}
    \includegraphics[width=0.49\linewidth]{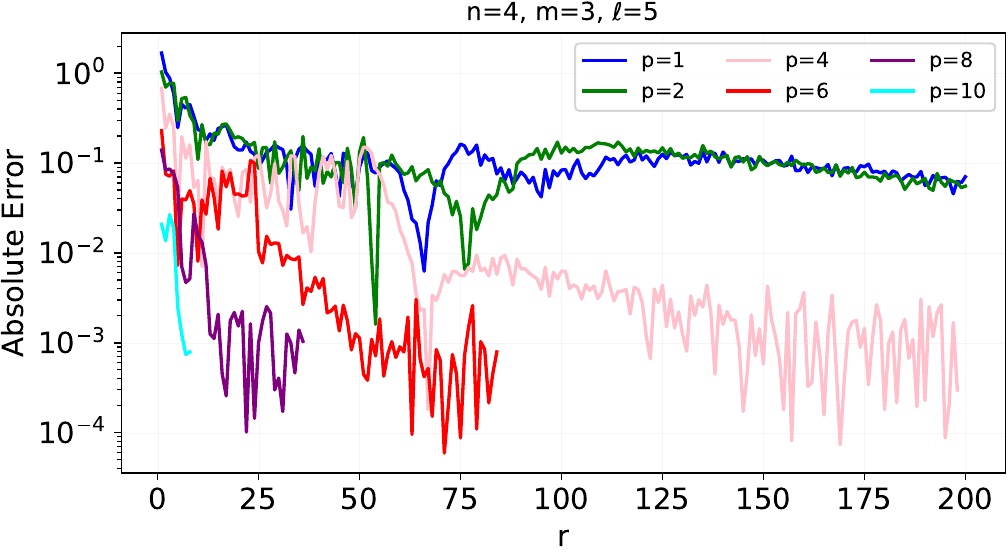}
    \caption{\textbf{Absolute error of the minimum eigenvalue estimate as a function of Trotter resources (order $p$ and steps $r$) for tQPE with a perfect initial state.} The absolute error is measured as the absolute difference between $E_0$ and the mean of the decoded eigenvalue estimates. If the Trotter error is sufficiently small then the entire distribution will be comprised of $\phi^*$ phases. Therefore, the absolute error rate is determined by both Trotter error, and the fundamental digitization error given by $\ell$ and $m$. The noise floor is contributed to by the finite sampling effect of $10^4$ shots measured, per parameter combination. Smaller $p$ results in substantially higher Trotter error, which translates to higher minimum eigenvalue absolute error if the QPE sampling has not converged to ``steady-state''. 
    Notice that when $\ell+m$ is larger, the required Trotter resources to get to the steady state noise floor are greater; this means that if the computation is constrained to use limited $p$ and $r$ trotterization, then making $\ell$ smaller is likely beneficial in terms of absolute error rate for sampling the minimum eigenvalue. This rule however is not always true; one of the other tradeoffs is what eigenvalue spacing does the initial state support, and to what resolution do we want to differentiate eigenvalues. If we do want very high accuracy small eigenvalue differentiation, particularly for large Hamiltonians, then this necessitates $\ell$ being very large and therefore also high trotterization resources. 
    Eq.~\eqref{equation:average_error} defines the absolute error measure. 
    }
    \label{fig:absolute_error_function_of_trotter_error_perfect_init}
\end{figure*}

\begin{figure*}[ht]
    \centering
    \includegraphics[width=0.49\linewidth]{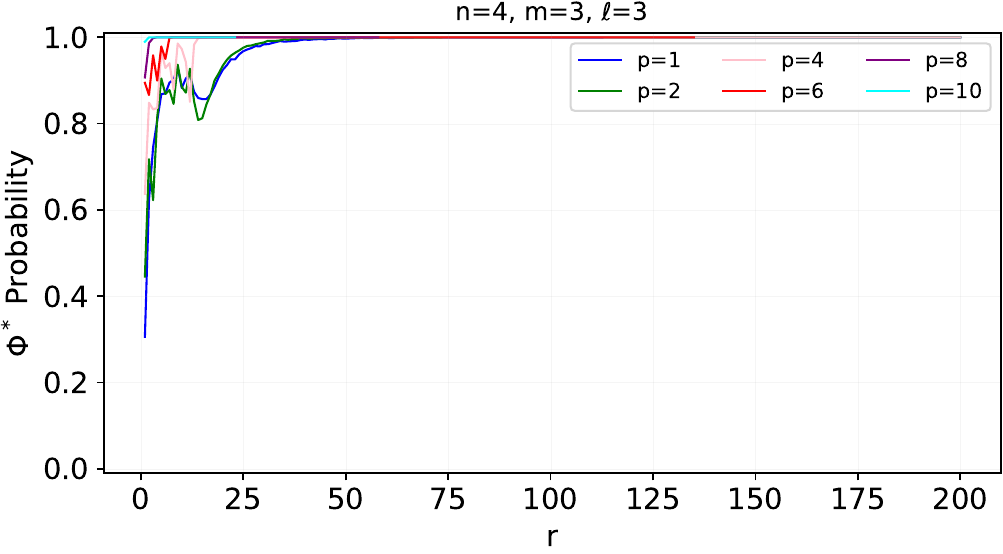}
    \includegraphics[width=0.49\linewidth]{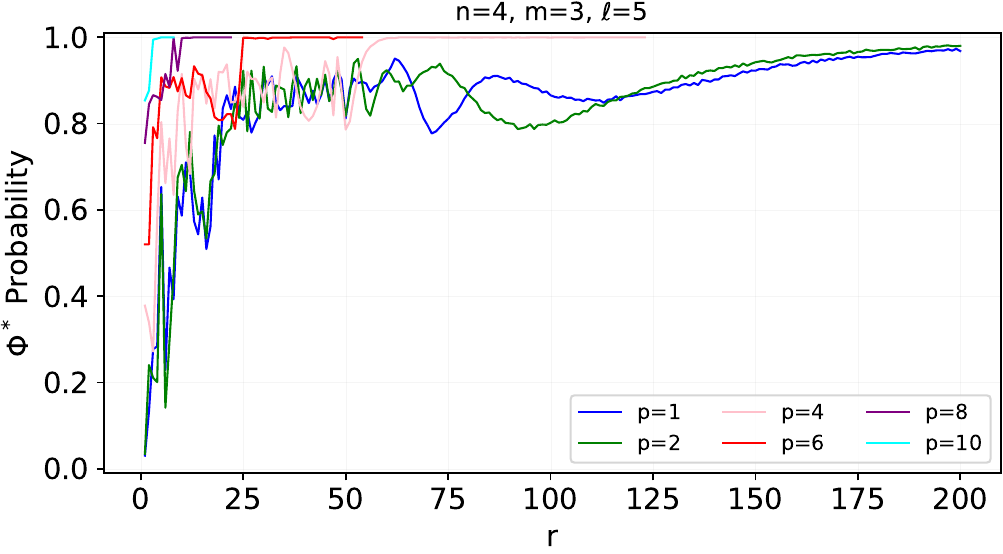}
    \caption{\textbf{Proportion of sampling phases in $\Phi^*$ as a function of Trotter resources (order $p$ and steps $r$) used in tQPE with a perfect initial state.} Because the initial state for the Hamiltonian evolution is perfect $\chi = 1$, which means that the sampling rate of $\Phi^*$ converges to $1$ as the Trotter error is decreased (within the sampling range of the finite shot count). Notably, the $\Phi^*$ sampling rate does not necessarily monotonically increase as a function of $r$, or $p$, as we may expect. 
    }
    \label{fig:optimal_phase_sampling_rate_perfect_init}
\end{figure*}

QPE, including tQPE, can fail to accurately separate different phases in the eigenspectrum of the quantum Hamiltonian being simulated due to several potential failure modes. The first is aliasing where the evolution time becomes too long and the phases wrap-around, resulting in ambiguity when the measured phases are translated back into eigenvalue estimates. Another is when the evolution time is not sufficiently long so as to use most of the available phase-space (e.g. all $2^{m+\ell}$ bit strings in the case of tQPE). If the evolution time is too small, then eigenphases will not be measured as distinct phase-bitstrings. This necessitates having a number of phase qubits (sometimes called ancilla qubits) that guarantees being able to differentiate two eigenvalues separated by some $\delta$. Here, the quantum Hamiltonians we are numerically simulating are all sufficiently small such that this differentiation is not an issue, but for larger QPE this is an important consideration. The perfect evolution time could be set by knowing the exact spectral range (difference between minimum and maximum eigenvalue), but this is very hard in general to know and in fact we want to use QPE to determine such spectral information to begin with. However, we do need to select a (base) evolution time that does not cause aliasing. To this end, we use 

\begin{equation}
    \label{equation:evolution_time_bound}
    t_0 = \frac{\pi}{\sum_{i \in \mathcal{H}} |c_i|}, 
\end{equation}

where $c_i \in \mathcal{H}$ is each coefficient in the Hamiltonian. This evolution time ensures that we do not exceed an evolution time corresponding to the spectral range of the quantum Hamiltonian. The use of this time evolution bound is to make the operation of tQPE realistic in the sense that when implementing standard QPE or tQPE we will only be able to use naive bounds on the spectral range in order to set the evolution time, since we will not know the exact spectral range.

\begin{figure*}[ht]
    \centering
    \includegraphics[width=0.49\linewidth]{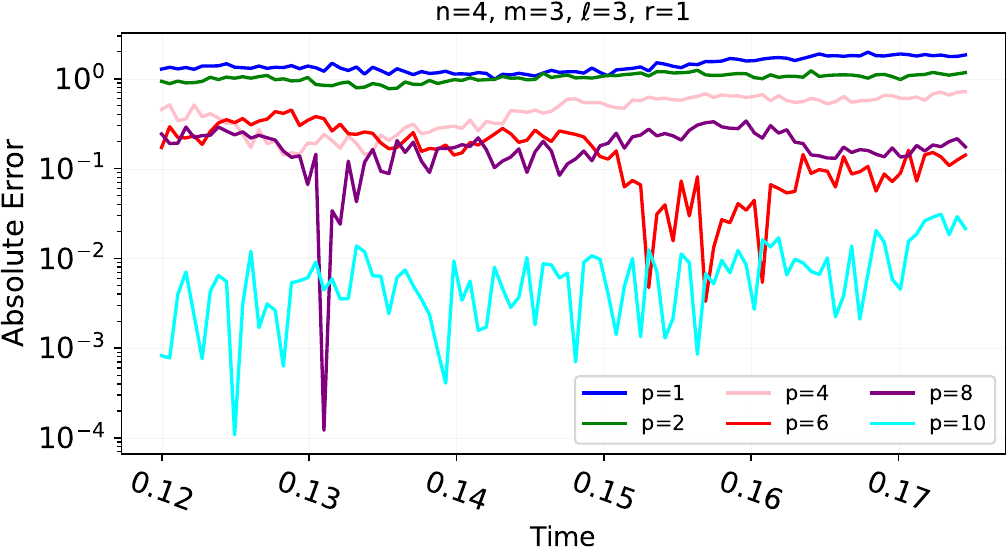}
    \includegraphics[width=0.49\linewidth]{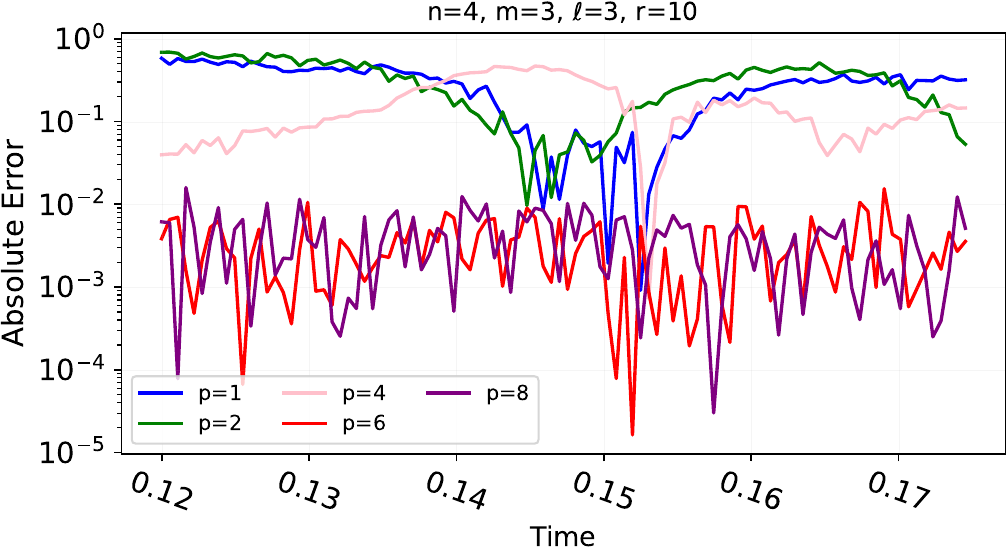}
    \includegraphics[width=0.49\linewidth]{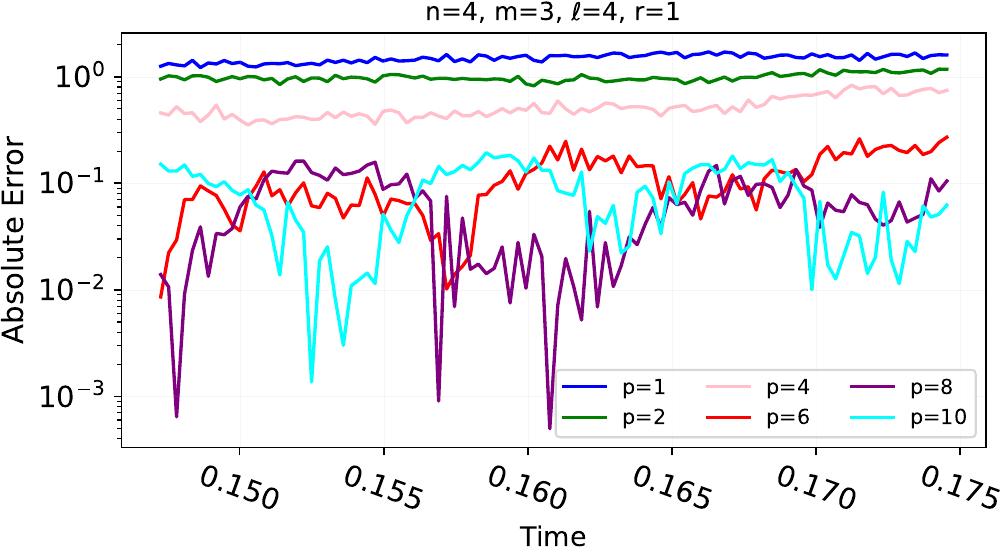}
    \includegraphics[width=0.49\linewidth]{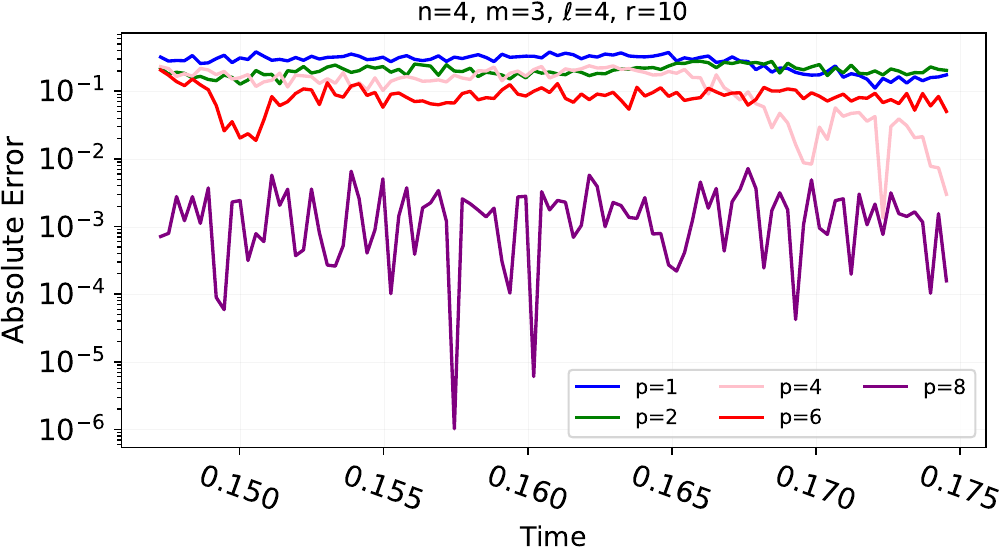}
    \caption{ \textbf{Absolute error from tQPE, with a perfect initial state, as a function of the (base) evolution time in $e^{i H t}$.} The absolute error is measured between the full eigenvalue estimate distribution, produced from a perfect initial state tQPE execution, and the true minimum eigenvalue $E_0$. 
    Note that the measured eigenvalue distribution has shot noise, which is the primary reason for the high-variation in error. Eq.~\eqref{equation:average_error} defines the absolute error measure. 
    }
    \label{fig:error_function_of_time_perfect_init}
\end{figure*}

\begin{figure*}[ht]
    \centering
    \includegraphics[width=0.49\linewidth]{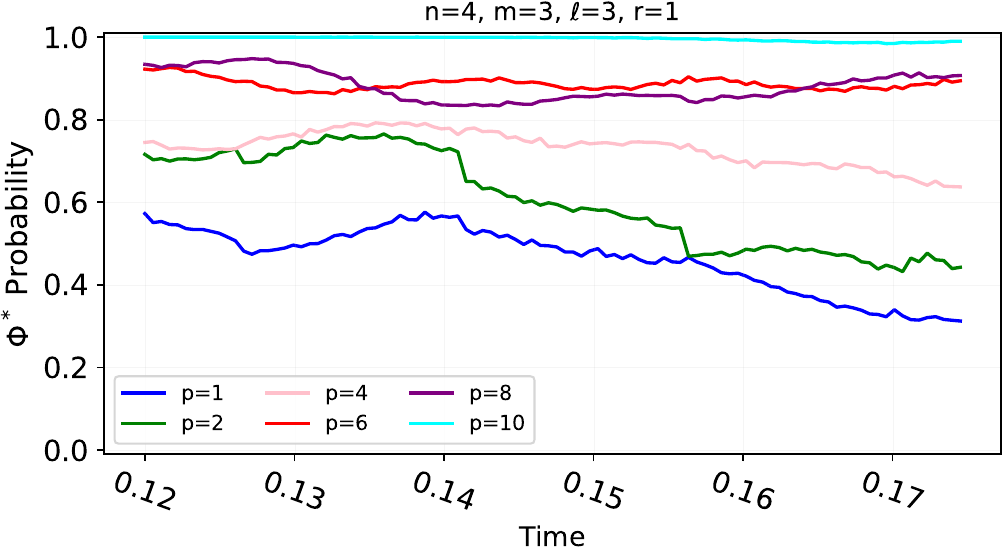}
    \includegraphics[width=0.49\linewidth]{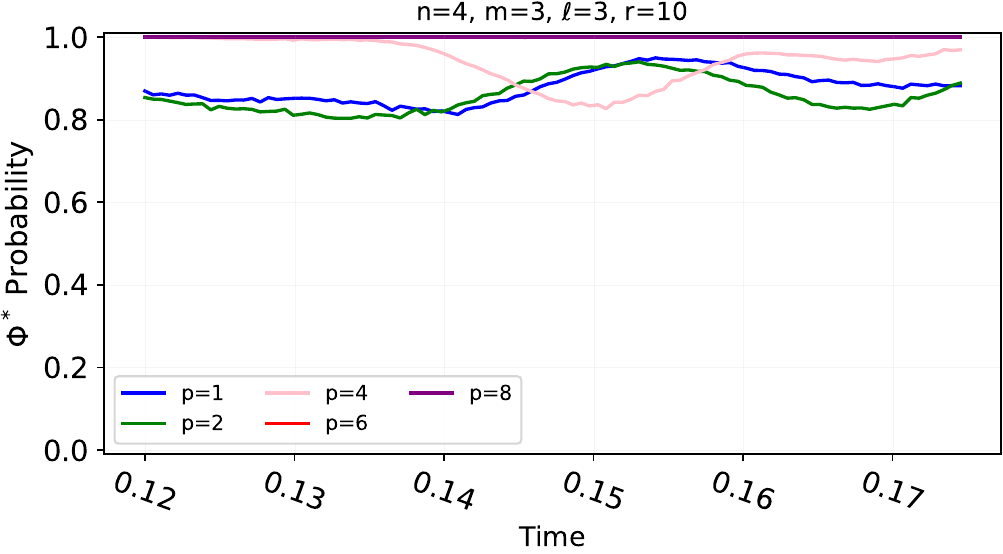}
    \includegraphics[width=0.49\linewidth]{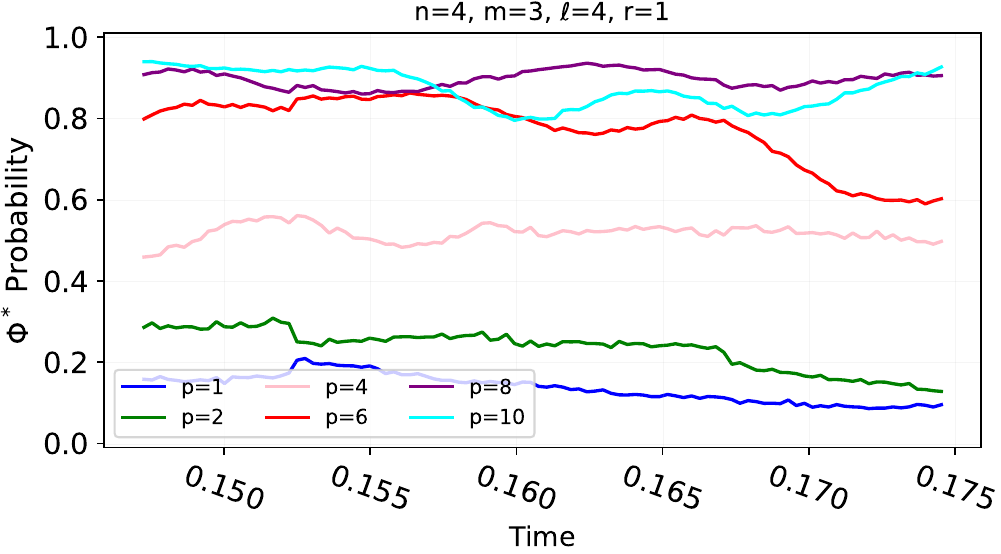}
    \includegraphics[width=0.49\linewidth]{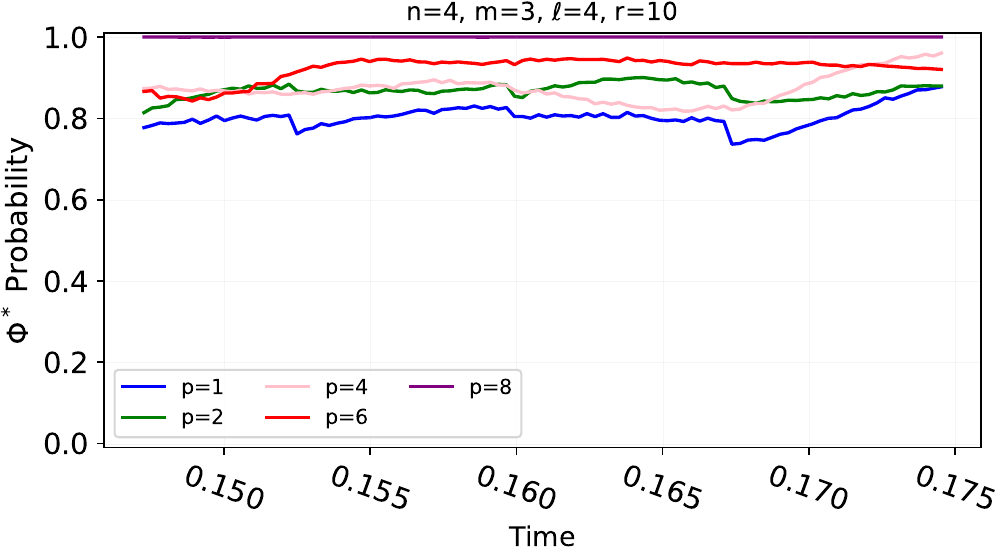}
    \includegraphics[width=0.49\linewidth]{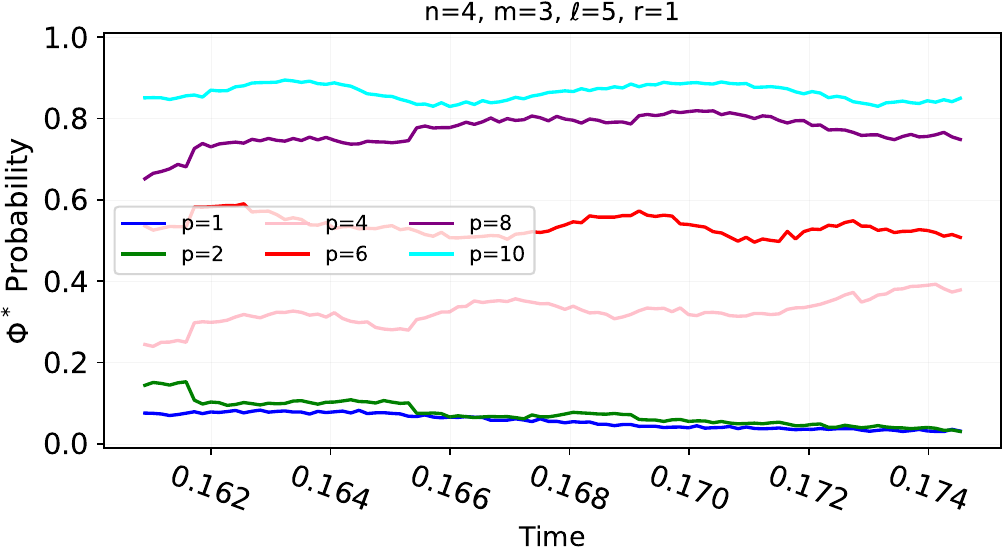}
    \includegraphics[width=0.49\linewidth]{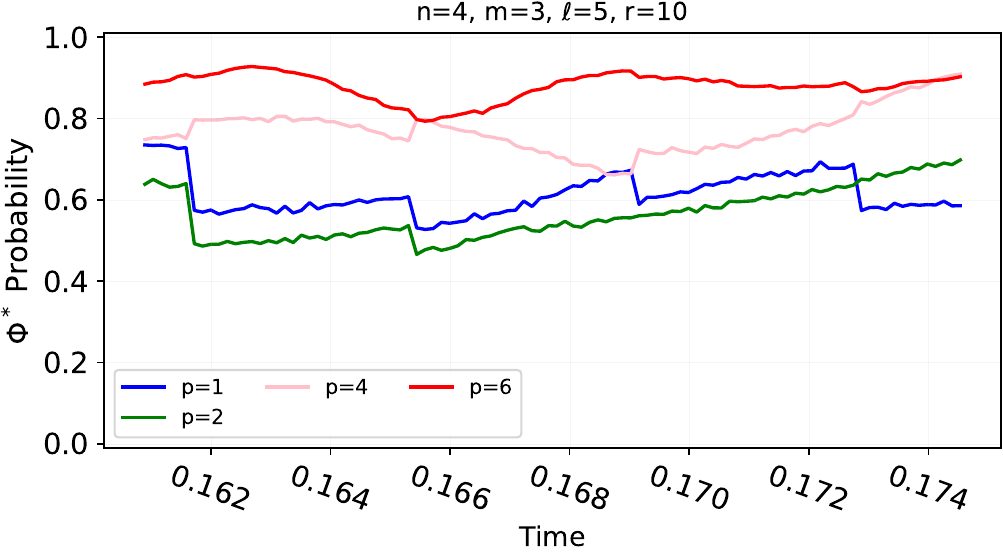}
    \caption{ \textbf{Sampling rate of $\Phi^*$ from tQPE with a perfect initial state as a function of the (base) evolution time in $e^{i H t}$.} When the Trotter order results in sufficiently low error (which is dependent on how large $\ell$ is), the measured $\Phi^*$ sampling rate reaches $1$ because the initial state is perfect. In general there is little dependence on the evolution time, unlike the strictly periodic dependence of standard QPE~\cite{pelofske2026numericalexperimentsparametersetting}. Interestingly, high Trotter error simulations, such as order $1$ and $2$ trotterization, do show some periodic time-dependence in the sampling rate of $\Phi^*$. The probabilities become small when $\ell$ is large, and the Trotter error is also large (due to $r$ and $p$ not being sufficiently large). 
    The set of optimal phases, $\Phi^*$, changes as a function of the evolution time; this means that the reported $\Phi^*$ probability is not necessarily the same when the evolution time changes. 
    }
    \label{fig:optimal_phase_sampling_rate_function_of_time_perfect_init}
\end{figure*}

\begin{figure*}[ht]
    \centering
    \includegraphics[width=0.49\linewidth]{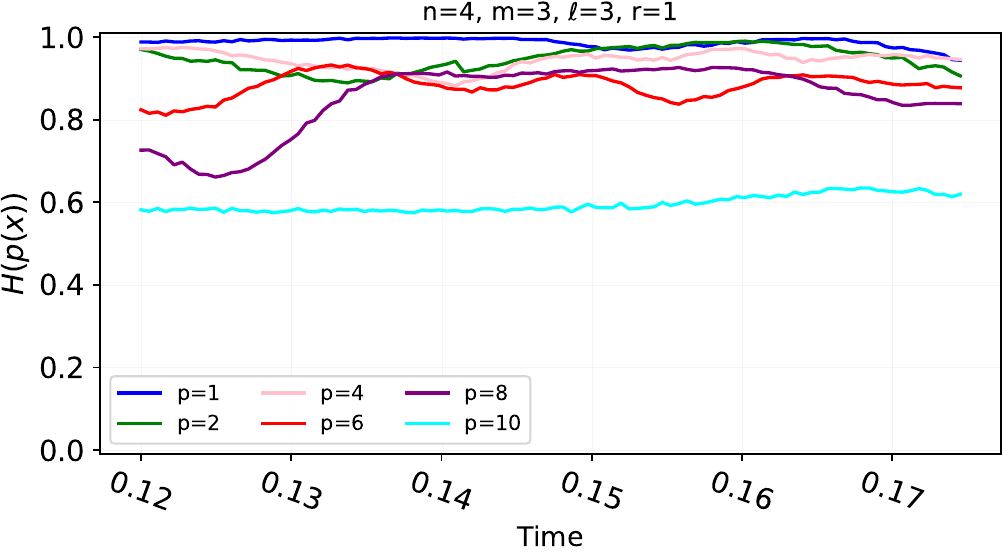}
    \includegraphics[width=0.49\linewidth]{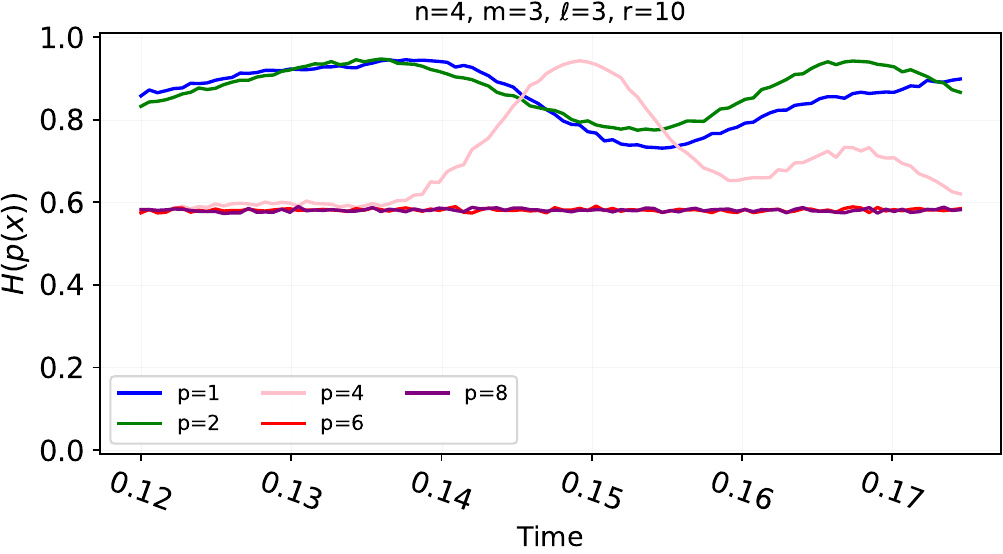}
    \includegraphics[width=0.49\linewidth]{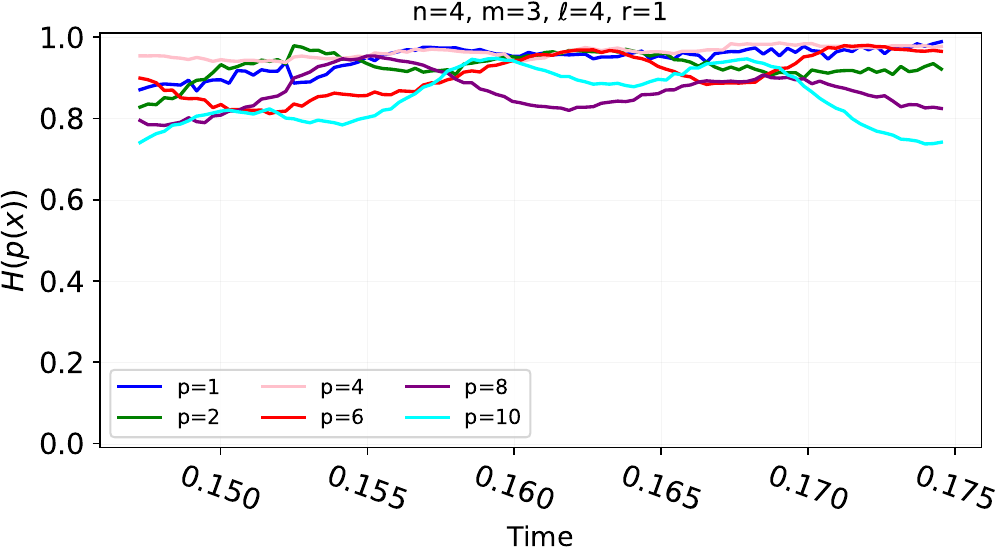}
    \includegraphics[width=0.49\linewidth]{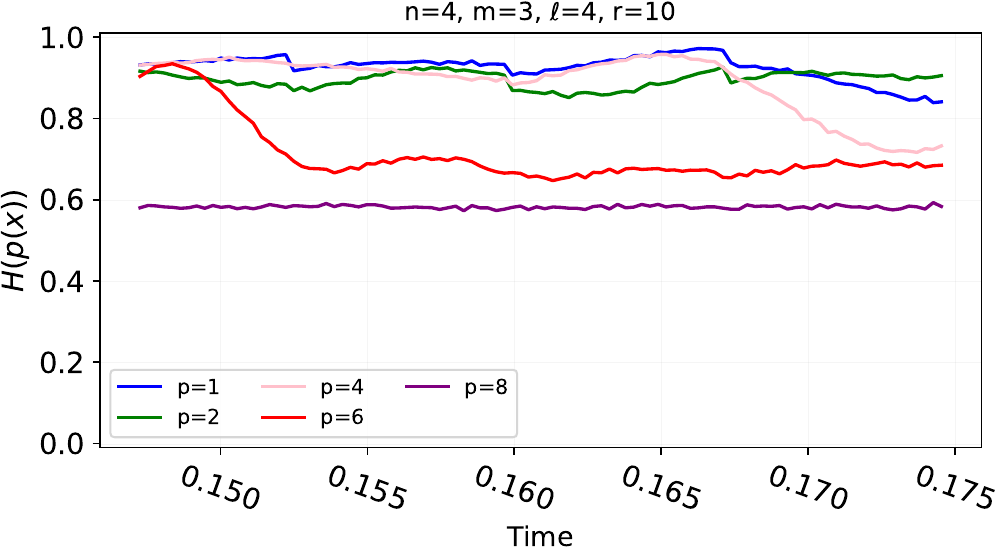}
    \includegraphics[width=0.49\linewidth]{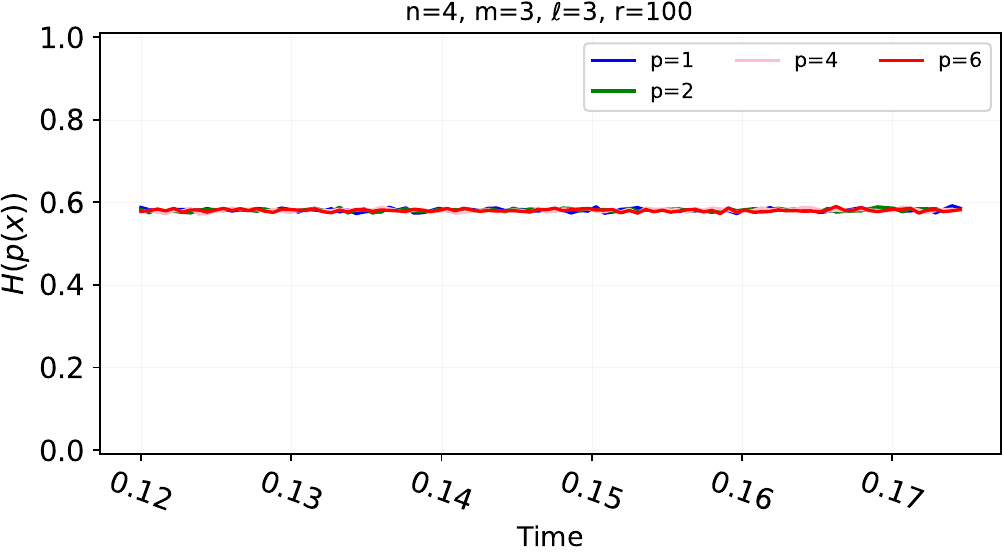}
    \includegraphics[width=0.49\linewidth]{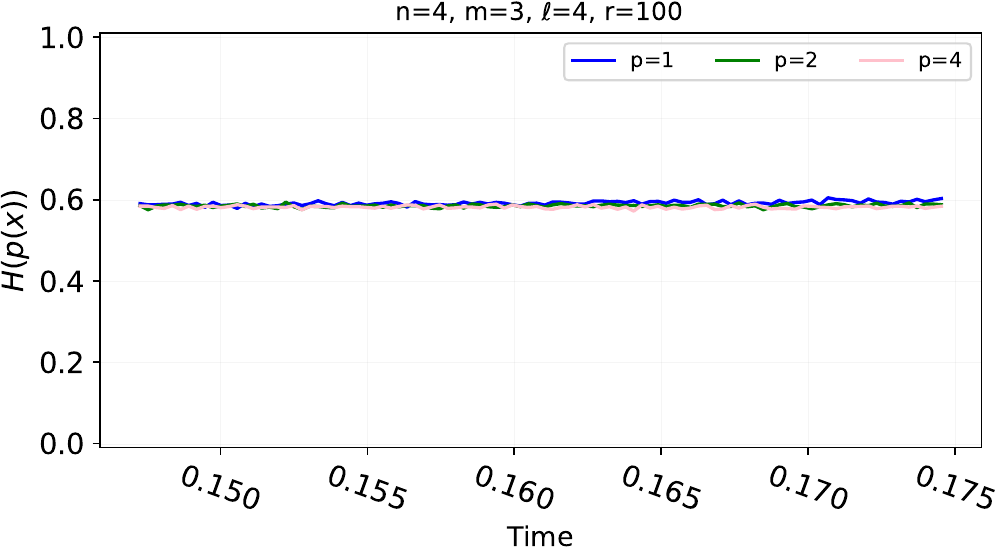}
    \includegraphics[width=0.49\linewidth]{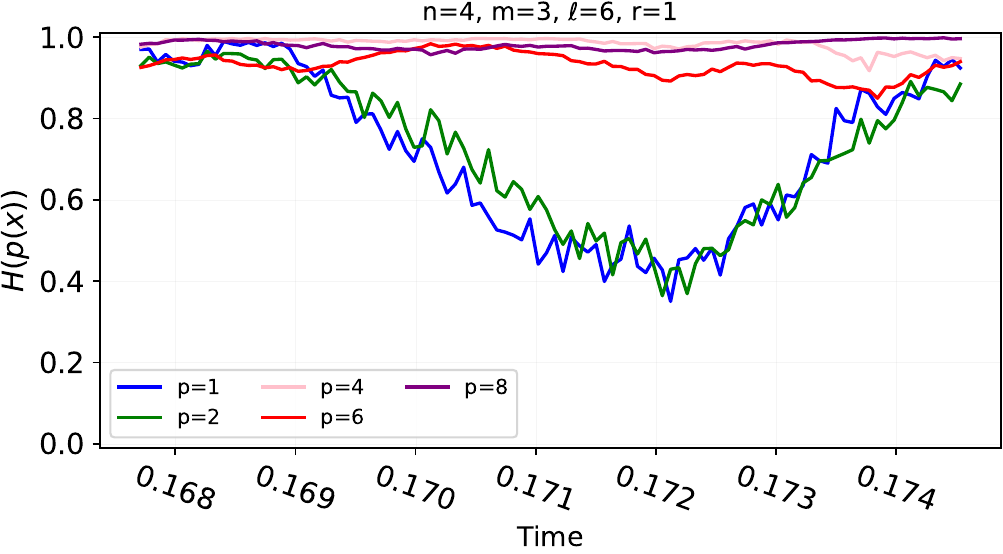}
    \includegraphics[width=0.49\linewidth]{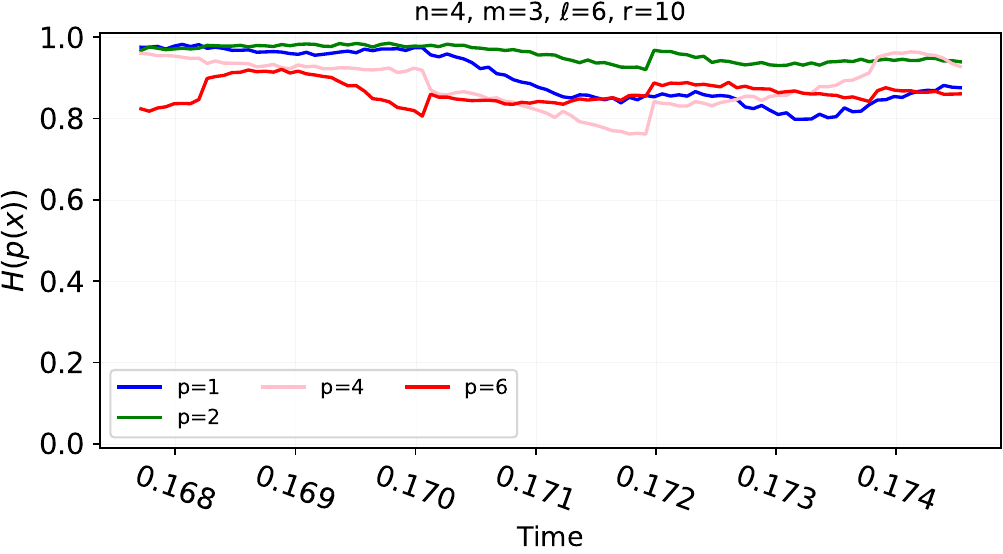}
    \caption{\textbf{Shannon entropy of the optimal tQPE phase distribution $\Phi^*$, with a perfect initial state. } Generally, there is not a dependence on the evolution time when tQPE has reached ``steady-state''. However, when the Trotter error is high, there is dependence in the form of rapid entropy changes, or gradual even periodic-type trends. Notice that the time-scale on the x-axes changes depending on $\ell$; this is scaled so that the evolution time switches between a handful of different phases, which in the case of the lower-right most plot we see several discontinuous jumps in the Shannon entropy corresponding to the boundary between different phases. 
    The two lower-middle plots show converged tQPE, where the Trotter error is sufficiently small (because $r=100$ and $\ell$ is relatively small), where the optimal phase sampling distribution no longer changes as the evolution time changes. 
    At this converged steady state the Shannon entropy is $\approx 0.58$ which means that the distribution is not entirely concentrated on one phase, but is also not uniform.  
    }
    \label{fig:shannon_entropy_optimal_phases_perfect_init}
\end{figure*}

\begin{figure*}[ht]
    \centering
    \includegraphics[width=0.49\linewidth]{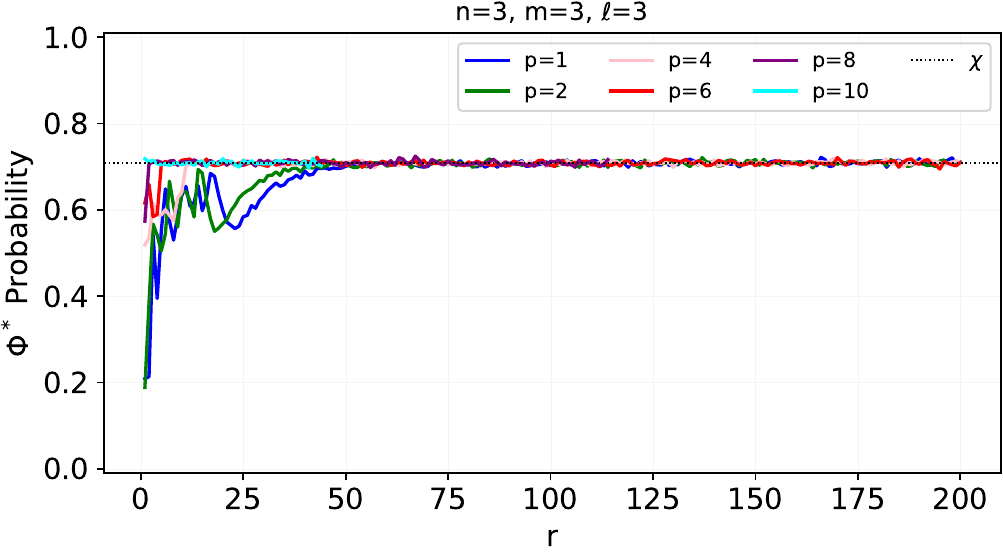}
    \includegraphics[width=0.49\linewidth]{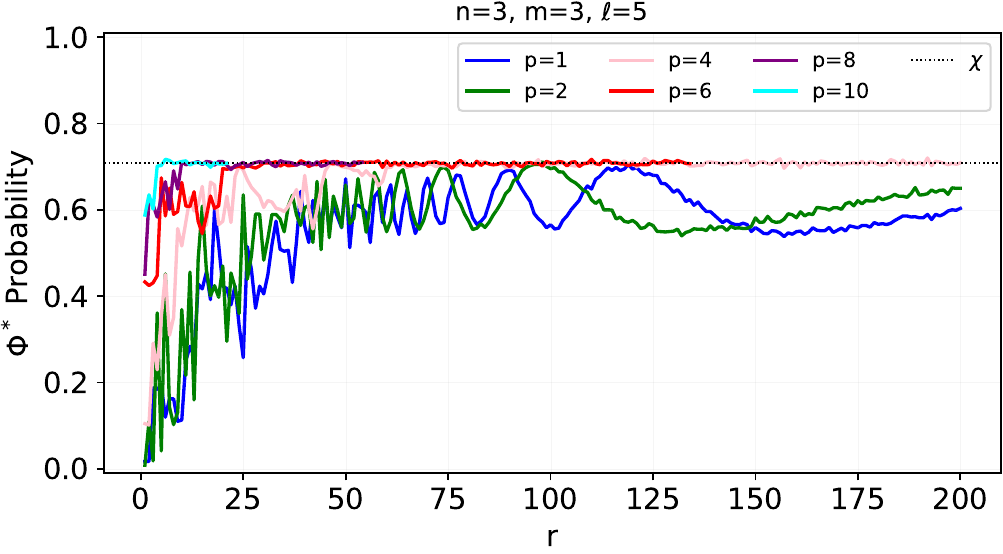}
    \includegraphics[width=0.49\linewidth]{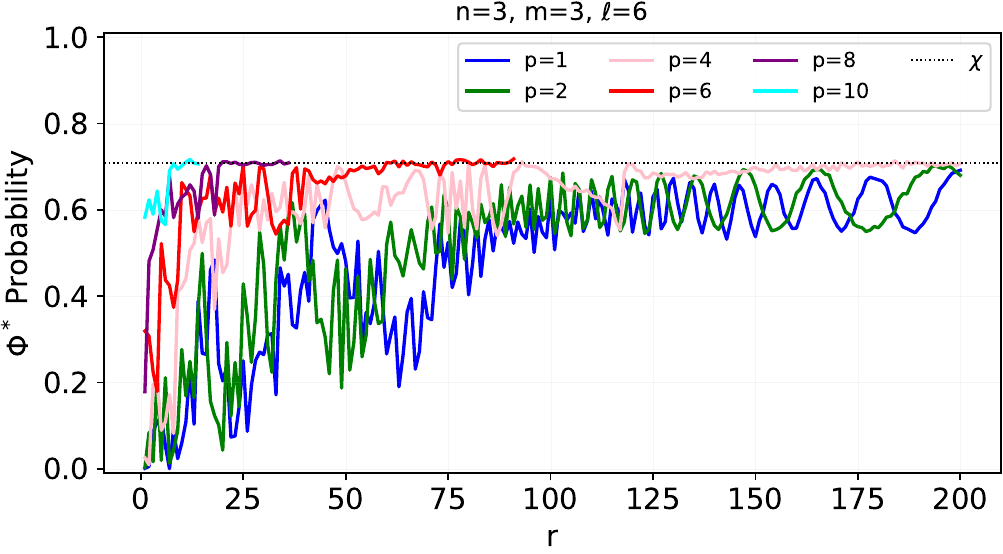}
    \includegraphics[width=0.49\linewidth]{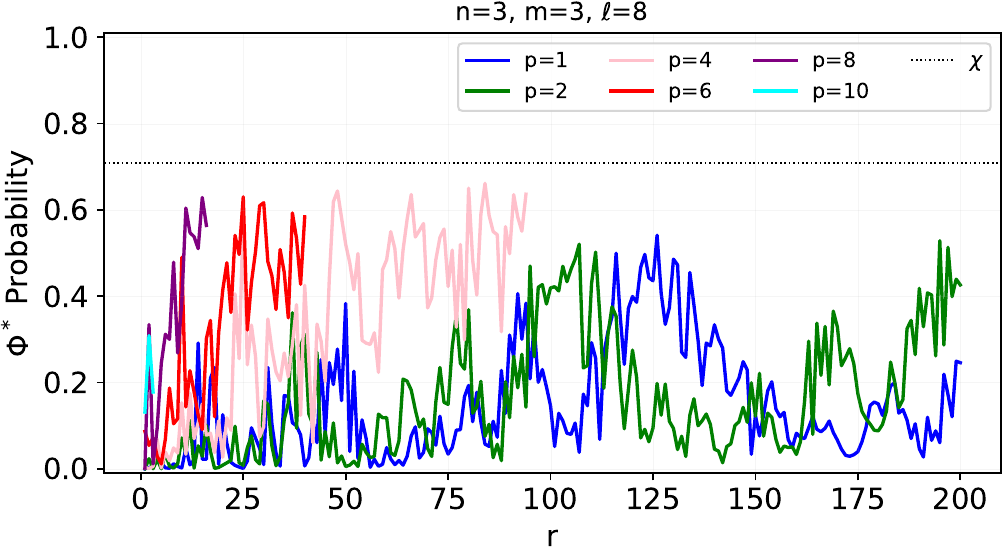}
    \caption{\textbf{tQPE optimal phase sampling probability with an imperfect initial state, whose overlap is $\chi \approx 0.708$. } At sufficiently low Trotter error, the optimal phase sampling rate converges to the overlap with the ground-state, given by the dashed black line $\chi$. For $\ell=3,5,6$ (top row, and bottom left) it is clear that by making $r$ and (or) $p$ sufficiently large, the optimal phase probability converges to $\chi$; larger $\ell$ means more Trotter resources are required to reach this ``steady-state''. However, for $\ell=8$, although the $\Phi^*$ probability tends to get larger as $p$ and $r$ increase, none of the trotterization schemes can reach the $\Phi^* \approx \chi$ steady-state. 
    }
    \label{fig:imperfect_initial_state_optimal_phase_sampling_rate}
\end{figure*}

In standard QPE, the distribution of the measured phases depends heavily on the evolution time used in the time dynamics~\cite{pelofske2026numericalexperimentsparametersetting, kaye2006introduction}. A natural question therefore is how the time evolution impacts tQPE. To this end, we vary the (base) evolution time to measure the $2K+1$ optimal phase bitstring sampling rate, the absolute error, and lastly the Shannon entropy~\cite{shannon1948mathematical} of the $2K+1$ phases so as to quantify distribution information about the optimal phase measurements of tQPE. The Shannon information entropy is defined, given the set of optimal taper QPE phases $\Phi^*$, let $p(x)$ denote the finite-shot sampling estimate of $\Phi^*$ from the Qiskit circuit executions, as 

\begin{equation}
    \label{equation:Shannon_entropy}
    H(p(x)) = -\sum_x p(x) \log_{2K+1} p(x), 
\end{equation}

where $x \in \Phi^*$, regardless of whether other phases are measured, and we normalize the measure by taking the log base $2K+1$ so that the maximum $H(p(x)$ is $1$. The Shannon entropy measure shows whether the phase distribution concentrates on a single phase readout ($H(p(x) = 0$), or is uniform over the span of the $2K+1$ phase bit measurements ($H(p(x) = 1$), or is somewhere in-between these two extremes. Shannon information entropy, although it has foundational meanings in information theory and thermodynamics, here we use it strictly as a canonical measure of a discrete probability distribution. It is a natural measure to use in order to quantify the distribution of $\Phi^*$, specifically to identify cases where the phase probabilities concentrate vs. are more uniformly spread over the $2K+1$ high-confidence region. For instance, a Shannon entropy close to zero could indicate that the phase probability distribution is concentrating on the digitized phase nearest to the true phase, thus producing a Dirac delta function $\delta$. However, the Shannon entropy may be close to $1$, which would correspond to a nearly uniform distribution and would show that tQPE is not concentrating probability onto the nearest digitized phase. Therefore, the Shannon entropy paired with histograms of the produced phase estimates will inform the characteristics of the distribution of $\Phi^*$, which is important to know when considering the digitization error of the computed phase. Moreover, Shannon entropy is a natural measure to determine sensitivity to change in evolution time. As the evolution time changes, the set of tQPE optimal phases $\Phi^*$ change as well, and Shannon entropy of that measured $\Phi^*$ distribution will be able to determine abrupt changes, as well as natural quantification of the steady-state distribution. 

The overlap between the ground-state and the initial state determines the overall success probability, both for QPE and tQPE. To this end, because the Heisenberg spin glass model can have degenerate ground-states, we measure the overlap, following ref.~\cite{pelofske2026numericalexperimentsparametersetting}, as

\begin{equation}
    \label{equation:degenerate_overlap}
    \chi = \sum_{\psi_2 \in v_0} \left| \langle \psi_1 \mid \psi_2 \rangle \right|^2, 
\end{equation}

where $\psi_1$ is the initial state prepared by $U_{\mathrm{prep}}^{\mathrm{state}}$, and $v_0$ is the set of minimum-eigenvalue orthonormal eigenvectors. Degenerate eigenvalues are determined by a precision of $1 \times 10^{-12}$. 

Finally, we describe the procedure for decoding measured phases into eigenvalue estimates. Each measured phase, which has $\ell+m$ bits, is translated into a normalized phase by $\phi_x = \frac{x}{2^{\ell+m}} \in [0, 1)$. Then the phase is shifted by 

$$
\bar{\phi}_x =
\begin{cases}
\phi_x, & \phi_x \le \frac{1}{2}, \\[2mm]
\phi_x - 1, & \phi_x > \frac{1}{2}.
\end{cases}
$$

Thus the eigenvalue estimate is

$$
\lambda_x = \frac{2 \pi}{t} \bar{\phi_x},
$$

where $t$ is the (base) evolution time. This means that phases greater than $\frac{1}{2}$ correspond to negative eigenvalues.

\begin{figure*}[htb]
    \centering
    \includegraphics[width=0.49\linewidth]{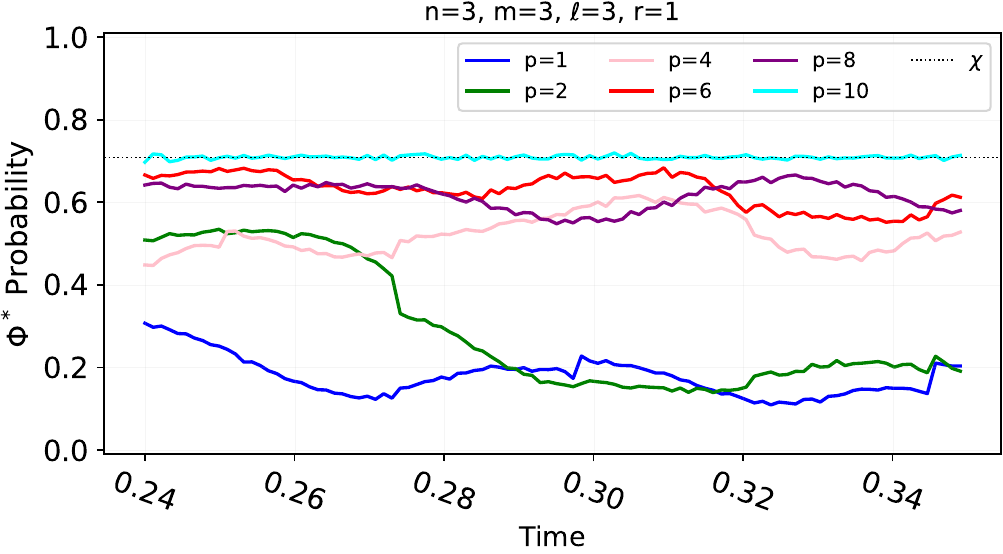}
    \includegraphics[width=0.49\linewidth]{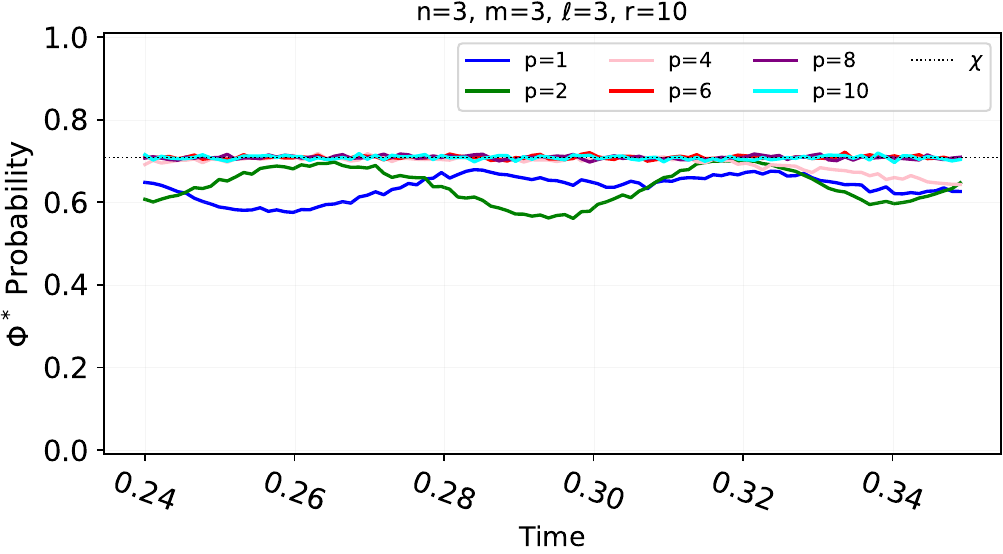}
    \includegraphics[width=0.49\linewidth]{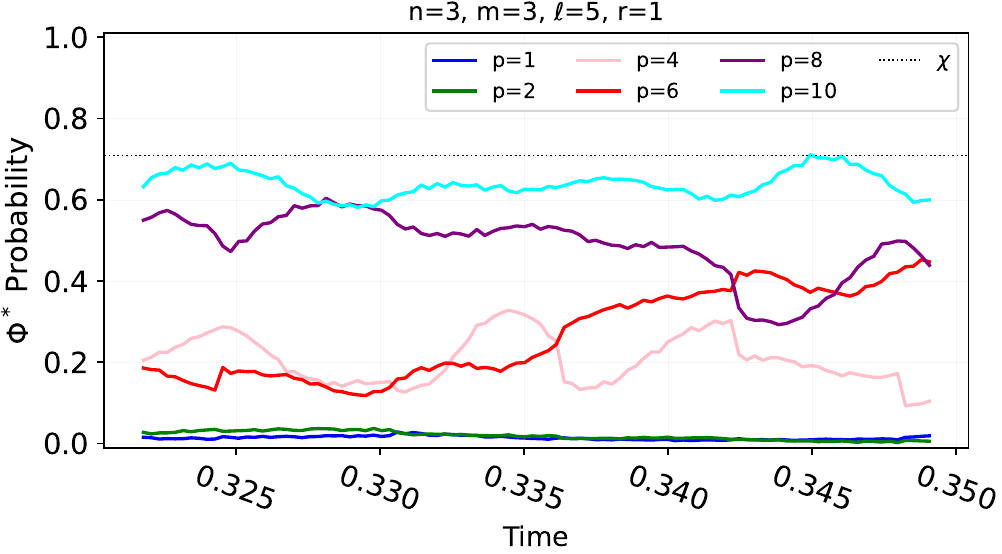}
    \includegraphics[width=0.49\linewidth]{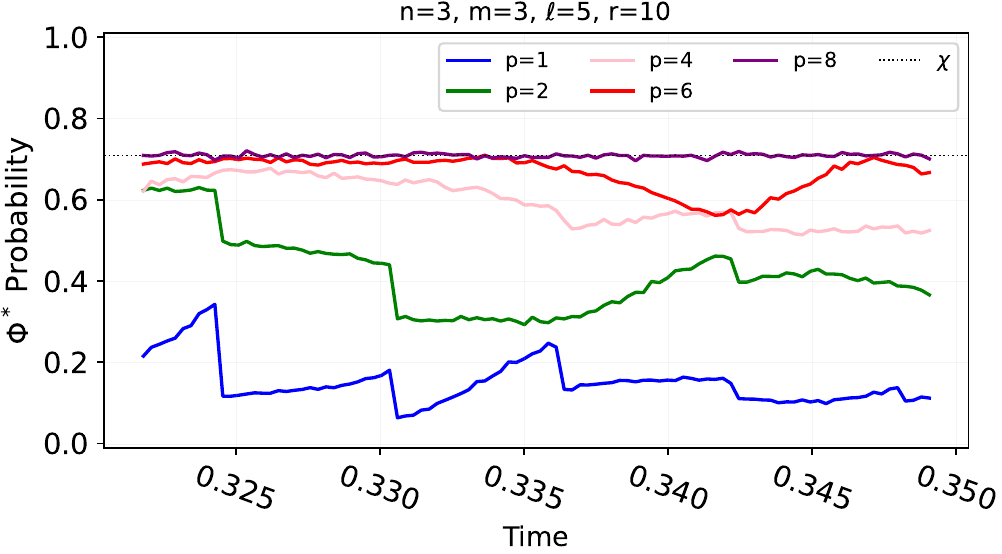}
    \caption{\textbf{tQPE optimal phase sampling probability with an imperfect initial state ($\chi \approx 0.708$) as the evolution time is changed.} The top row reports results for $\ell=3$, the bottom row reports results for $\ell=5$. 
    There is some time dependence on the evolution time if the Trotter error is sufficiently high such that the sampling has not converged to $\chi$. 
    The top-left plot shows that only $p=10,r=1$ Trotterization causes the tQPE sampling to converge to $\Phi^* \approx \chi$, whereas the top-right plot shows that if we increase to $r=10$ then $p=4,6,8,10$ all converge to the steady-state sampling where there is no apparent dependence on evolution time. By comparison, increasing the bits of precision to $\ell=5$, only $p=8,r=10$ converges to the steady-state (and $p=10$ was not able to be executed). 
    }
    \label{fig:imperfect_initial_state_vary_time}
\end{figure*}

%%%%%%%%%%%%%%%%%%%%%%%%%%%%%%%%%%%%%%%%%%%%%%%%
\section{Results}\label{section:Results}
%%%%%%%%%%%%%%%%%%%%%%%%%%%%%%%%%%%%%%%%%%%%%%%%

Here we systematically evaluate the effects of Trotter error, initial state overlap, evolution time, and the total bit-precision of the phase estimate. 

First, we examine the effect of Trotter error, where the Hamiltonian simulation initial state $U_{\mathrm{prep}}^{\mathrm{state}}$ is an exact minimum-eigenvalue eigenvector, which means the initial state is perfect. Fig.~\ref{fig:eigenvalue_distributions_perfect_init} reports distributions of the eigenvalue estimates with both high ($p=r=1$) and low ($p=10, r=1$) Trotter error. This demonstrates that sufficiently high Trotter error can result in substantial non-physical eigenvalue estimates. Specifically, these eigenvalue estimates are outside of the spectral range of the target quantum Hamiltonian $\mathcal{H}$; the reason is because of the approximate unitary that is actually implemented for each phase qubit $\Tilde{U}_j$. tQPE inherently suppresses spectral leakage due to phase-grid misalignment, however here the results show a different physical mechanism of a type of spectral leakage due to Trotter error. This makes the eigenvalue estimate non-physical for the original intended Hamiltonian. Moreover, there exist intermediate Trotter error rates which still result in a clear concentration of probabilities near a good estimate of the minimum eigenvalue $E_0$, while there is also a distribution of ``side-lobes''.

Fig.~\ref{fig:absolute_error_function_of_trotter_error_perfect_init} reports the absolute error of the minimum eigenvalue estimate as the Trotter error is decreased (by making both $r$ and $p$ larger), where the initial state is perfect. The absolute error is defined as

\begin{equation}
    \label{equation:average_error}
    | E_0 - \frac{\sum_i \lambda_i}{n_\text{samples}} |,
\end{equation}

where each $\lambda_i$ is a sampled (and then decoded) eigenvalue estimate. In Fig.~\ref{fig:absolute_error_function_of_trotter_error_perfect_init}, the error reaches a noise floor due to digitization error and due to finite sampling. Because the overlap is perfect ($\chi=1$), if the Trotter error is sufficiently low then the sampling rate of $\Phi^*$ goes to $1$, which is shown in Fig.~\ref{fig:optimal_phase_sampling_rate_perfect_init}. In many cases, especially when $p$ is large, there is incomplete data for all $r$; this occurs because on a single compute node the circuit is too large either to be defined or to be simulated using the Qiskit statevector simulator due to out of memory errors.

Fig.~\ref{fig:optimal_phase_sampling_rate_perfect_init} shows one of the key features of tQPE, which is that at sufficiently small Trotter error the sampling of $\Phi^*$ reaches a ``steady-state'', which specifically means that the sampling has converged to the overlap value. The caveat is that this steady-state here is approximate because of three reasons. First, finite sampling, second there is still a finite amount of Trotter present in the circuit, and finally the taper state is approximate and so there can be a very small rate of spectral leakage. Therefore, we will refer to this convergence as the tQPE ``steady-state'', where the sampled probability of $\Phi^* \approx \chi$. Note that standard QPE also has this characteristic of a steady-state phase probability~\cite{pelofske2026numericalexperimentsparametersetting} at some threshold of finite Trotter error. Where and why this steady-state property occurs deserves further future investigation, but what it means for tQPE is that the Trotterization that is used in the circuit does not need to be arbitrarily expensive to reach the sampling predicted by the initial state overlap, indeed in Fig.~\ref{fig:optimal_phase_sampling_rate_perfect_init} convergence occurs with relatively few Trotter resources (although, this is of course dependent on $\ell$).

Fig.~\ref{fig:absolute_error_function_of_trotter_error_perfect_init} illustrates an important tradeoff in tQPE sampling, which applies in general to QPE accuracy; digitization error vs Trotter error (the error source of the initial state has been removed here because the initial state is perfect). Although increasing $\ell$ increases the digitization accuracy of the minimum eigenvalue estimate, this comes at the cost of more Trotter resources that are needed in order to reach that higher-accuracy eigenvalue sampling. This means that if the quantum resources that are available to approximate the Hamiltonian time-evolution unitaries have high error rates, then making $\ell$ small will result in lower overall error rates. Fig.~\ref{fig:absolute_error_function_of_trotter_error_perfect_init} shows this for instance with Trotter order $2$ and $r$ of $200$. When $\ell=3$, the error rate on the sampled eigenvalue distribution is between $10^{-3}$ and $10^{-2}$, whereas when $\ell=5$ the error rate is just below $10^{-1}$. This means that for second order Trotterization, at $r=200$, it is better to use $\ell=3$ compared to a higher precision phase readout such as $\ell=5$. tQPE has many examples of these types of tradeoff curves, which are all important considerations for future practical implementations of tQPE.

Next, we examine the time-dependence of tQPE. We vary the evolution time from $t_0$ (see Eq.~\eqref{equation:evolution_time_bound}) to $t_0 - \frac{t_0 \cdot 20}{2^{m+\ell}}$ with $100$ linearly spaced times. This choice of evolution time allows the evolution time to move the QPE eigenphase across several different grid-points (phase digitizations), where the resolution of that interval is set by the total number of phase qubits. First, Fig.~\ref{fig:error_function_of_time_perfect_init} reports the absolute error as a function of the evolution time. This shows that there is some time dependence, but not in the form of a clear trend, instead local trends emerge when the Trotter error is high. Moreover, as expected when the evolution time was fixed, the sampling rate of $\Phi^*$ also converges to $1$ if the Trotter error is sufficiently small, regardless of the (base) evolution time as shown in Fig.~\ref{fig:optimal_phase_sampling_rate_function_of_time_perfect_init}. 
Finally, Fig.~\ref{fig:shannon_entropy_optimal_phases_perfect_init} reports the $\Phi^*$ Shannon entropy as Trotter resources ($p$ and $r$) are increased which shows that the steady-state entropy when the error is sufficiently small approaches $\approx 0.58$ regardless of $\ell$, for $m=3$. However, when Trotter error is large $H(p(x))$ can vary from $0.4$ to $1$ where $1$ is a uniform distribution over $\Phi^*$. Comparing to Fig.~\ref{fig:eigenvalue_distributions_perfect_init}, we can see that the steady-state Shannon entropy corresponds to a large concentration on the phases closest to the physical minimum eigenvalue phase, a substantial probability dropoff for the other nearby phases.

Lastly, we examine initial states of the Hamiltonian evolution ($U_{\mathrm{prep}}^{\mathrm{state}}$ in Fig.~\ref{fig:tapered_QPE_circuit}) that do not have perfect overlap with the ground-state. Here we focus on a $3$-qubit Heisenberg spin-glass quantum Hamiltonian, where the initial state used is the graph state~\cite{PhysRevA.69.062311, hein2006entanglementgraphstatesapplications} on the $3$ qubits, which has overlap $\chi \approx 0.708$. The (clique) graph state is defined as 

\begin{equation}
    \label{equation:graph_state}
    |G\rangle = \left( \prod_{(i,j) \in E} \mathrm{CZ}_{ij} \right) H^{\otimes n} |0^n \rangle,
\end{equation}

where $E$ is the clique edgeset. The choice of a graph state as an initial state here is arbitrary; it serves as an example of a relatively easy-to-prepare state, which happens to have high overlap with this particular $3$-qubit quantum Hamiltonian. Moreover, it is not the case that graph states tend to have high ground-state overlap for these Heisenberg models~\cite{pelofske2026numericalexperimentsparametersetting}; the overlap decreases at larger model sizes. 
Fig.~\ref{fig:imperfect_initial_state_optimal_phase_sampling_rate} reports the $\Phi^*$ probability as more Trotter resources (Trotter order and Trotter steps) are used, which shows that the $\Phi^*$ probability converges to $\chi$ as expected (within the spectral leakage error rate given by the approximate taper state, and finite sampling). In this way, the overlap is the steady-state sampling rate of $\Phi^*$ for tQPE. Importantly, this is only true if there are sufficiently many $\ell$ phase qubits such that the gap between the ground state and the first excited state can be differentiated into different phase bins -- if $\ell$ is small and the first excited state and the ground-state are too close then the observed phase probability will be greater than $\chi$ corresponding to the ground-states. Interestingly, the $\Phi^*$ probability does not necessarily monotonically improve as more Trotter steps are used to approximate the unitary Hamiltonian time evolution. 
Next, Fig.~\ref{fig:imperfect_initial_state_vary_time} reports $\Phi^*$ as the tQPE evolution time is changed, which shows that at steady-state there is no observed dependence on the evolution time, in stark contrast to standard QPE~\cite{pelofske2026numericalexperimentsparametersetting}. However, when the Trotter error is high (for instance, $r=1$ for first and second order Trotter in Fig.~\ref{fig:imperfect_initial_state_vary_time}-left), there is strong dependence on the evolution time. This was also true when the initial state was the ground state eigenvector (Fig.~\ref{fig:optimal_phase_sampling_rate_function_of_time_perfect_init}). However, the high-Trotter error dependence on the time evolution does not have a clear trend, or periodicity. 

The numerical calculations we have presented show clear non-monotonic trends as the Trotter resources $p$ and $r$ are increased. In part, this is because Trotter error does not necessarily decrease as $p$, or $r$, are increased; ref.~\cite{pelofske2026numericalexperimentsparametersetting} showed this using direct numerical calculations varying both $r$ (Trotter steps) and $p$ (Trotter order).

%%%%%%%%%%%%%%%%%%%%%%%%%%%%%%%%%%%%%%%%%%%%%%%%
\section{Conclusion}\label{section:Conclusion}
%%%%%%%%%%%%%%%%%%%%%%%%%%%%%%%%%%%%%%%%%%%%%%%%

We have presented a full circuit level implementation of tapered QPE. Importantly, this implementation allows us to examine several key aspects of the practical considerations of tQPE, including non-physical eigenvalue estimates that arise due to poor approximation of $e^{i H t}$ which are highlighted by Fig.~\ref{fig:eigenvalue_distributions_perfect_init}-(top row). Moreover, if $\ell$ is small, even if $m$ is sufficiently large ($m=3$ is sufficiently large), the digitization error on the minimum eigenvalue estimate $E_0$ can be relatively large because of the tQPE high confidence phase region, which for $m=3$ consists of $7$ digitized phases.

In the numerical results that we report, we focus on the task of minimum eigenvalue computation. However, focusing on measuring phases that correspond to the minimum eigenvalue is a very stringent requirement for QPE. In general, sampling from the low-energy eigenvalue spectrum, not strictly the smallest eigenvalue, of a quantum Hamiltonian could be sufficient. We focus on the minimum eigenvalue because it is the hardest task, but also highly application relevant for computational physics and computational chemistry.

Like in standard QPE, the Trotter error does not need to be arbitrarily small in order to reach the steady-state optimal phase sampling rate; Figs.~\ref{fig:imperfect_initial_state_optimal_phase_sampling_rate} and  \ref{fig:optimal_phase_sampling_rate_perfect_init} illustrate this property. For tQPE the steady-state optimal phase sampling probability is determined strictly by the overlap quantity (Eq.~\eqref{equation:degenerate_overlap}). An interesting future research direction is to quantify when this steady-state phase sampling occurs; this property occurs for standard QPE as well~\cite{pelofske2026numericalexperimentsparametersetting}.

We focus on numerical computations of only two Heisenberg spin glass models, with at most $4$ qubits. These are very small system sizes relative to the regime where we would want QPE to be executed. This system size is necessitated by the nature of the numerical computations; these circuits are extremely large and require HPC resources to execute in any meaningful amount of time. Moreover, we focus on only single instances because the focus of this study is to highlight the different error sources within tQPE. These algorithmic properties of tQPE are independent of system size -- system size impacts the performance of tQPE substantially but this is because of i) the high cost of good initial state preparation and ii) the high cost of approximating the controlled $e^{i H t}$ operations. Moreover, $\ell$ (and potentially $m$) must be increased when the system size increases because the initial state overlap will likely necessitate high accuracy differentiation of low-energy eigenstates in the energy spectrum. Increasing $\ell$ then makes the required Trotter resources even greater. In summary, larger system size implications are very important for tQPE in terms of the overall circuit complexity, however, here we focus on the phase sampling characteristics of tQPE on small instances.

Another future research direction is applying phase randomization~\cite{Patel_2026, Linden_2022, lu2022unbiasedquantumphaseestimation, PhysRevLett.129.030503}, which we did not do for this tQPE implementation. 

Finally, we have shown numerically that although the low-Trotter-error tQPE computations have no strong dependence on the Hamiltonian evolution time, if there is Trotter error present then there is some dependence on the evolution time, albeit with no clear periodicity or trend direction. The overall finding is that Trotter error can substantially change the properties of tQPE, from error rates on the minimum eigenvalue sampling, to time-dependence, to the distribution of phases, but not in a clearly predictable way.

%%%%%%%%%%%%%%%%%%%%%%%%%%%%%%%%%%%%%%%%%%
\section*{Acknowledgments}\label{sec:acknowledgments}
%%%%%%%%%%%%%%%%%%%%%%%%%%%%%%%%%%%%%%%%%%
We thank Yigit Subasi, David Raveh, and Andrew Sornborger for productive discussions on taper QPE. We thank Yigit Subasi for providing the taper state preparation vectors. 
Los Alamos National Laboratory is operated by Triad National Security, LLC, for the National Nuclear Security Administration of the U.S. Department of Energy (Contract No. 89233218CNA000001). This material is based upon work supported by the U.S. Department of Energy, Office of Science, National Quantum Information Science Research Centers, Quantum Science Center.
This research used resources provided by the Los Alamos National Laboratory Institutional Computing Program. 
LANL report LA-UR-26-27837.

\appendix

\section{Taper State Numerical Calculation Details}
\label{section:appendix_taper_states}

This section briefly describes the numerical calculations used to numerically compute the approximate taper state, specifically the statevector \texttt{tPREP} in Fig.~\ref{fig:taper_state_vector_circuit}. The numerically calculated vector (with rounded coefficients) for $\ell = 3$, $m=3$, is

\[
\begin{bmatrix}
0.7233+0.0i \\
-0.4731-0.0232i \\
0.1179+0.0116i \\
-0.0059-0.0009i \\
0.0+0.0i \\
-0.0059+0.0009i \\
0.1179-0.0116i \\
-0.4731+0.0232i 
\end{bmatrix}.
\]

This is the state labeled as \texttt{tPREP} in Fig.~\ref{fig:taper_state_vector_circuit}. The calculation of this vector \texttt{tPREP} involves a numerical computation heuristic developed in Ref.~\cite{Patel_2026} that is a type of optimized bandwidth-limited DPSS taper state preparation, which ultimately is bottlenecked by an exact diagonalization procedure, where the size of the matrix to be diagonalized scales with $m$. However, $m \leq 5$ gives more than sufficient error suppression, which overall makes the procedure relatively efficient, although practically making $\ell$ large also increases the computation time. The optimization of the bandwidth-limited DPSS taper state targets the number of optimal phases $\Phi^*$ to be $2^m-1$. Specifically, the procedure generates a vector of size $2^m -1$, and then the single zero entry is padded in the middle of the vector; this vector is a statevector object which is used to initialize the circuit. $\ell$ does not contribute to the size of this vector, however the entries are changed slightly when $\ell$ changes. In Ref.~\cite{Patel_2026} it was observed that fixing \texttt{tPREP} for some reasonable $\ell$ results in reasonable performance of the taper state. 

The numerically calculated vector (with rounded coefficients) for $\ell = 10$, $m=3$, is

\[
\begin{bmatrix}
0.7226+0.0i\\
-0.474 -0.0002i\\
0.1191+0.0001i\\
-0.0061-0.0i\\
0.0 + 0.0i\\
-0.0061+0.0i\\
0.1191-0.0001i\\
-0.474 +0.0002i
\end{bmatrix}.
\]

\section{Heisenberg Model Values}
\label{section:appendix_model_values}

Here we give the exact model weights of the arbitrarily chosen quantum Hamiltonians used in the numerical calculations in this study. 

The 3 qubit model instance weights are $J_{0, 1}^x + J_{0, 2}^x -  J_{1, 2}^x - J_{0, 1}^y - J_{0, 2}^y - J_{1, 2}^y + J_{0, 1}^z - J_{0, 2}^z -  J_{1, 2}^z$.

The 4 qubit model instance weights are $-J_{0, 1}^x - J_{0, 2}^x + J_{0, 3}^x + J_{1, 2}^x + J_{1, 3}^x - J_{2, 3}^x +
- J_{0, 1}^y - J_{0, 2}^y + J_{0, 3}^y - J_{1, 2}^y - J_{1, 3}^y - J_{2, 3}^y + 
J_{0, 1}^z - J_{0, 2}^z - J_{0, 3}^z + J_{1, 2}^z + J_{1, 3}^z - J_{2, 3}^z$.